\documentclass[%
 reprint,
showpacs,
 amsmath,amssymb,
 amsfonts,
 aps,
 pra,
showkeys
]{revtex4-1}

\usepackage{graphicx}% Include figure files
\usepackage{dcolumn}% Align table columns on decimal point
\RequirePackage{atbegshi}
\usepackage{hyperref}

\usepackage{bbold}
\usepackage{verbatim}

\usepackage{subfigure}

\usepackage{units}
\usepackage[]{todonotes}

\usepackage[utf8]{inputenc}
\usepackage[T1]{fontenc}
\usepackage[ngerman,english]{babel}
\usepackage{slashed}
\usepackage{upgreek}
\usepackage{dsfont}

\usepackage{hyperref}
\hypersetup{
    colorlinks=false,
    citecolor=black,
    filecolor=black,
    linkcolor=black,
    urlcolor=black,
    linktocpage
}

\newcommand{\Tr}{\operatorname{Tr}}
\newcommand{\TrC}{\ensuremath{\Tr_c}}
\newcommand{\real}{\operatorname{Re}}

\newcommand{\ee}{\ensuremath{\textrm{e}}}
\newcommand{\ii}{\ensuremath{\textrm{i}}}

\newcommand{\NGroup}{\ensuremath{\text{N}}}
\newcommand{\NSigma}{\ensuremath{\text{N}_\sigma}}
\newcommand{\NTau}{\ensuremath{\text{N}_\tau}}

\newcommand{\Nc}{\ensuremath{\text{\NGroup}_\text{c}}}
\newcommand{\Nf}{\ensuremath{\text{\NGroup}_\text{f}}}

\newcommand{\VSpatial}{\ensuremath{V}}

\newcommand{\Loewe}{LOEWE-CSC}
\newcommand{\Sanam}{SANAM}

\newcommand{\Juqueen}{JUQUEEN}

\newcommand{\Qcd}{QCD}
\newcommand{\LQcd}{LQCD}

\newcommand{\clqcd}{CL\kern-.25em\textsuperscript{2}QCD}
\newcommand{\codename}{\clqcd}

\newcommand{\psibar}{\bar{\psi}} 
\newcommand{\chiralcond}{\ensuremath{\langle \psibar \psi \rangle}}

\newcommand{\mpi}{\ensuremath{m_{\pi}}}

\newcommand{\MFermion}{\ensuremath{D}}

\newcommand{\MFermionMinus}{\ensuremath{\MFermion^{-1}}}

\newcommand{\Action}{\ensuremath{\mathcal S}}

\newcommand{\SFermion}{\ensuremath{\Action_{\text{f}}}}

\newcommand{\SGauge}{\ensuremath{\Action_{\text{gauge}}}}

\newcommand{\G}{\ensuremath{g}}
\newcommand{\GSq}{\ensuremath{\G^2}}

\newcommand{\Center}{\ensuremath{\mathbb Z}}

\newcommand{\ZNc}{\ensuremath{\Center(\Nc)}}

\newcommand{\Link}{\ensuremath{U}}

\newcommand{\LinkDag}{\ensuremath{\Link^{\dagger}}}

\newcommand{\LatSpacing}{\ensuremath{a}}
\newcommand{\LatMass}{\ensuremath{m}}

\newcommand{\LatMassWilson}{\ensuremath{\kappa}}

\newcommand{\LatMassWilsonTric}{\ensuremath{\LatMassWilson^\text{tric}}}
\newcommand{\LatMassWilsonTricHeavy}{\ensuremath{\LatMassWilsonTric_\text{heavy}}}
\newcommand{\LatMassWilsonTricLight}{\ensuremath{\LatMassWilsonTric_\text{light}}}

\newcommand{\Plaq}{\ensuremath{P_{\mu\nu}}}

\newcommand{\LatX}{\ensuremath{n}}
\newcommand{\LatXSpatial}{\ensuremath{\mathbf{\LatX}}}
\newcommand{\LatY}{\ensuremath{m}}

\newcommand{\LatCoupling}{\ensuremath{\beta}}
\newcommand{\LatCouplingC}{\ensuremath{\LatCoupling_c}}

\newcommand{\Poly}{\ensuremath{L}}
\newcommand{\PolyAbs}{\ensuremath{|\Poly|}}
\newcommand{\PolyIm}{\ensuremath{\Poly_\text{Im}}}

\newcommand{\PolyPhase}{\ensuremath{\phi}}
\newcommand{\PolyMod}{\ensuremath{\hat{\Poly}}}
\newcommand{\PolyModAbs}{\ensuremath{|\PolyMod|}}
\newcommand{\PolyModPhase}{\ensuremath{\varphi}}

\newcommand{\PolyImExp}{\ensuremath{\langle \PolyIm \rangle }}

\newcommand{\Binder}{\ensuremath{B_4}}
\newcommand{\Temp}{\ensuremath{T}}

\newcommand{\Tc}{\ensuremath{\Temp_c}}
\newcommand{\Mu}{\ensuremath{\mu}}
\newcommand{\MuI}{\ensuremath{\mu_i}}
\newcommand{\MuIc}{\ensuremath{\MuI^c}}

\newcommand{\GenObs}{\ensuremath{X}}

\newcommand{\Susc}{\ensuremath{\chi}}

\begin{document}

\preprint{PREPRINT???}

\title{The nature of the Roberge-Weiss transition in \texorpdfstring{$N_f=2$}{Nf=2} QCD with Wilson fermions}

\author{Owe Philipsen}
 \email{philipsen@th.physik.uni-frankfurt.de}
\author{Christopher Pinke}%
 \email{pinke@th.physik.uni-frankfurt.de}
\affiliation{
 Institut f\"{u}r Theoretische Physik - Johann Wolfgang Goethe-Universit\"{a}t, Germany\\
 Max-von-Laue-Str. 1, 60438 Frankfurt am Main
}

\date{\today}

\begin{abstract}
At imaginary values of the quark chemical potential $\mu$, Quantum Chromodynamics shows an interesting phase structure due to an exact center, or Roberge-Weiss (RW), symmetry.
This can be used to constrain QCD at real $\mu$, where the sign problem prevents Monte Carlo simulations of the lattice theory.
In previous studies of this region with staggered fermions it was found that the RW endpoint, where the center transition changes from first-order to a crossover, depends non-trivially on the quark mass: for high and low masses, it is a triple point connecting to the deconfinement and chiral transitions, respectively, changing to a second-order endpoint for intermediate mass values.
These parameter regions are separated by tricritical points.
Here we present a confirmation of these findings using Wilson fermions on $N_\tau=4$ lattices.
In addition, our results provide a successful quantitative check for a heavy quark effective lattice theory at finite density. 
\end{abstract}

\pacs{12.38.Gc, 05.70.Fh, 11.15.Ha}
\keywords{QCD phase diagram}
\maketitle

\section{Introduction}

The \Qcd\ phase diagram at finite temperature \Temp\ and chemical potential 
\Mu\ is currently under investigation both theoretically and experimentally, 
with a particular interest in the search for
a potential critical endpoint (CEP). Because of its non-perturbative
nature at the energy scales of interest,
the only theoretical method to access \Qcd\ without truncations
is via simulations of its discretized version, Lattice \Qcd\ (\LQcd).
At zero quark chemical potential \Mu, 
the nature of the thermal \Qcd\ transition for $N_f=2+1$ flavors depends
on the quark mass configuration.
For degenerate infinitely heavy or massless quarks, 
there are first-order deconfinement and chiral phase transitions, respectively, at  some critical temperatures \Tc.
In the vicinity of these limits, there are regions of first-order transitions 
which are separated by $Z(2)$ second-order lines from a crossover region, 
where the physical point is located \cite{Aoki:2006we,deForcrand:2006pv}.
The nature of the transition in the $\Nf=2$ chiral limit is not settled yet.
A recent review of the phase diagram from the lattice 
is provided in \cite{Philipsen:2011zx}.

At fi\-ni\-te real chemi\-cal pot\-ential, the fermion de\-ter\-min\-ant 
becomes com\-plex.
This so-called sign problem prevents simulations using importance 
sampling.
By contrast, at purely imaginary values of the chemical potential, 
$\mu=i\mu_i, \mu_i\in \mathbb{R}$,
there is no sign problem and standard simulation techniques 
can be applied. In particular, the critical lines separating the first-order
from the crossover regions continue as critical surfaces to imaginary
chemical potential and terminate in tricritical lines at $\mu_i=\pi\Temp/3 $
\cite{deForcrand:2010he,Bonati:2010gi}.
Their location constrains the phase diagram at zero and real \Mu\ 
and in particular explains the negative curvature of the chiral 
critical surface at $\Mu=0$ obtained previously \cite{deForcrand:2006pv,deForcrand:2008vr}.

So far, \LQcd\ studies at imaginary chemical potential have been carried out predominantly 
using the staggered fermion discretization, 
investigations with Wilson fermions have only been 
started recently \cite{Nagata:2011yf,Nagata:2012pc,Wu:2013bfa,Alexandru:2013uaa}.
An independent confirmation of the phase structure found in 
\cite{deForcrand:2010he,Bonati:2010gi}
with a different discretization is of high interest because of
potential problems with the rooting
of staggered fermions \cite{Sharpe:2006re}.
Furthermore, a three-dimensional effective theory of \LQcd\ 
based on the hopping expansion of Wilson fermions has been put forward recently, which
allows to simulate heavy quarks at all chemical potentials
\cite{Fromm:2011qi}. 
The full \LQcd\ results presented here 
provide a successful check of the predictive power of the effective
theory.

\section{The Roberge-Weiss symmetry}

\begin{figure*}
        \centering
\subfigure{
\includegraphics{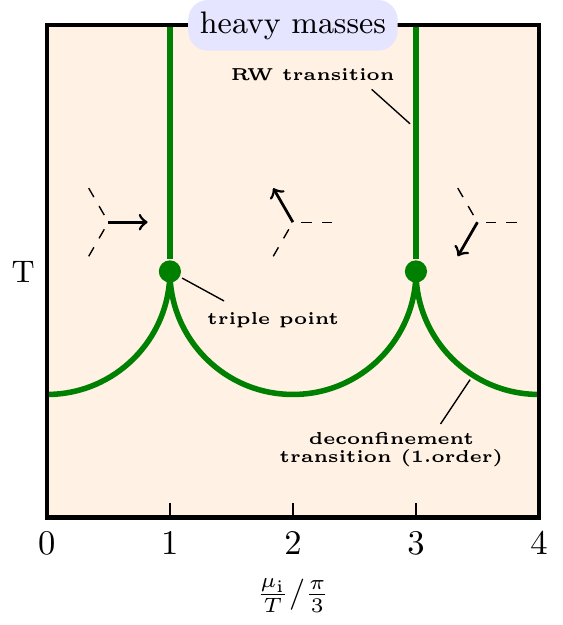}}
~\includegraphics{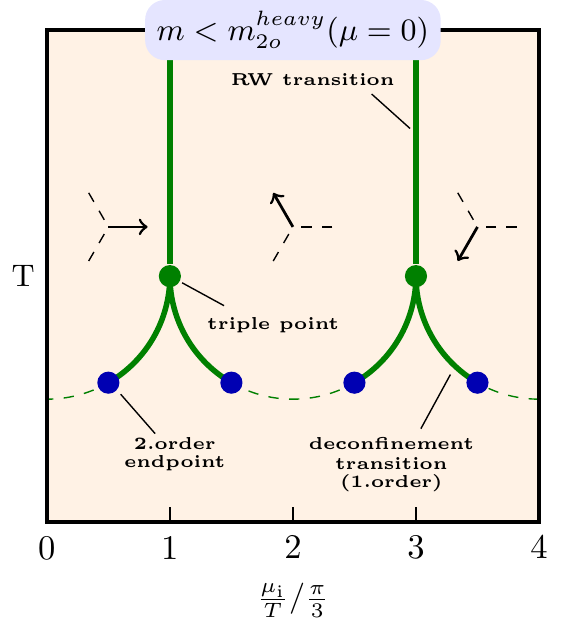}\\
\includegraphics{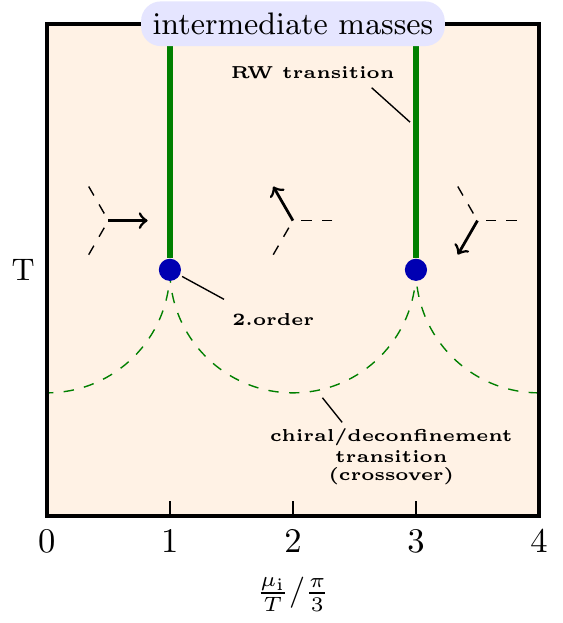}
~\includegraphics{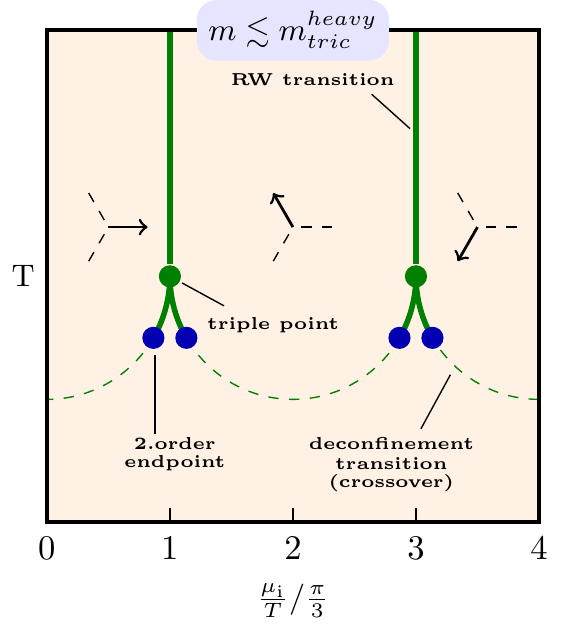}
        \caption{
					Schematic phase diagram of QCD at imaginary chemical potential. 
					The solid vertical lines show the first-order RW-transitions at \MuIc\ between the different \ZNc\ sectors, which are 
characterized by the phase of the Polyakov loop.
					Below $T_c$, the RW transitions are crossover.
					Beginning in the top left corner, the quark mass is decreased clockwise, changing the nature of the deconfinement and chiral phase transitions as indicated.
					}
				\label{fig:rwpd}
\end{figure*}

\begin{figure}
 		\centering
   \includegraphics[scale=0.8]{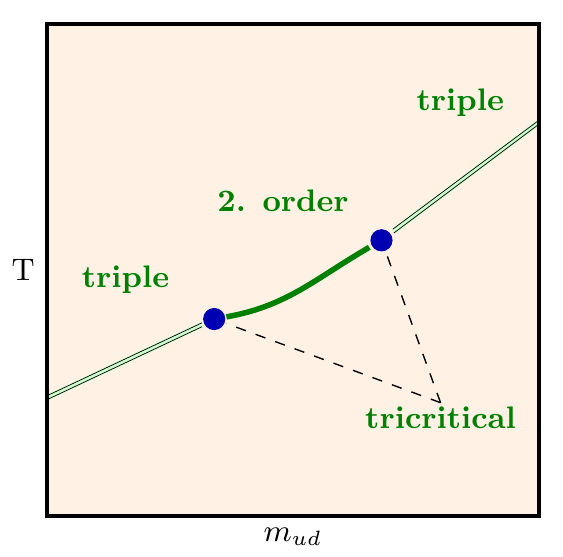}
 		\caption{RW endpoint as function of mass (schematic).}
 		\label{fig:rw_endpoint}
\end{figure}

The grand canonical partition function of QCD for arbitrary quark masses and 
at finite chemical potential
is even under charge conjugation. Moreover it
is invariant under non-trivial global center transformations of the gauge 
group, provided the quark chemical potential is shifted by a center element.
These exact symmetries read
\begin{eqnarray}
	\label{ZPeriodicityMui}
	Z(\mu)&=&Z(-\mu),\\ 
Z \left(\Mu \right)&=&Z \left (\Mu + 2\pi\ii k/\Nc\ \right), 
k \in \mathbb N\;.
\end{eqnarray} 
The periodicity in the imaginary chemical potential is called 
Roberge-Weiss (RW) symmetry  \cite{Roberge:1986mm}, for an elementary
introduction see \cite{Philipsen:2010gj}.
The general phase structure due to these symmetries 
was worked out in \cite{Roberge:1986mm} and is shown in Figure \ref{fig:rwpd}.
For the critical values 
\begin{equation}
	\label{MuICr}
	\MuIc =   (2k + 1)\ \pi\Temp/\Nc  \;\;, k=0,1\ldots\Nc-1\;,
\end{equation}
there are transitions between adjacent \ZNc\ sectors of the theory.
A particular center sector can be identified by the phase of 
the Polyakov loop $\Poly = \PolyAbs \ee^{\ii \PolyPhase}$.
However, due to the periodicity of the partition function 
all physical observables are invariant under shifts $\mu_i+ 2\pi k/\Nc$. 
At low temperatures the transition between adjacent sectors 
is a crossover whereas it is a first-order phase transition 
at high temperatures \cite{Roberge:1986mm,deForcrand:2002ci,D'Elia:2002gd}.
Consequently, the first-order Roberge-Weiss transition has an endpoint.
On the other hand, the deconfinement and chiral transition lines
extend into the \MuI\ region and meet the first-order RW line in the 
RW endpoint \cite{deForcrand:2010he,Bonati:2010gi}.
Therefore the nature of this end- or meeting point is non-trivial and depends on 
\Nf\ and the fermion mass.
This is sketched in Figure \ref{fig:rwpd}. For large masses, 
the deconfinement transition at $\mu=0$ is of first-order and 
joins the RW endpoint, which  
is a triple point.
As the mass is lowered, the $\mu=0$ transition passes through the second-order line and becomes a crossover.
This carries over to the \MuI\ region where the second-order point approaches 
the RW endpoint from $\mu=0$ with decreasing mass.
The remaining first-order line 
shrinks until it eventually meets the RW point.
At this mass value one has a tricritical point.
The same happens when coming from the chiral limit, increasing the mass, 
at least for $\Nf=2,3$ \cite{deForcrand:2010he,Bonati:2010gi}.
For fixed flavor content and for $\mu_i=\MuIc$, 
there is then a phase diagram
as in Figure \ref{fig:rw_endpoint}, showing the nature
of the RW endpoint as a function of quark mass.
%Varying the quark masses, a picture similar to the situation at $\Mu=0$ emerges naturally: For low and high masses there are regions of triple points, which are bounded from a second-order region by tricritical lines \cite{Bonati:2012pe}.
An order parameter can be defined by introducing the modified Polyakov loop $\PolyMod = \Poly \ee^{\ii \theta} = \PolyModAbs \ee^{\ii \PolyModPhase}$.
Its phase \PolyModPhase\ indicates the \ZNc\ sector the system is currently in and its average takes on the values zero and $k ( 2\pi/\Nc), k = 0,\ldots, \Nc-1$, for the low and high temperature phases, respectively.
% \begin{equation}
% 	\label{PolyakovLoopModPhaseExp}
% \langle \PolyModPhase \rangle = 
% % \begin{cases}
% % 0\;\text{ confined phase} \\
% % \;\text{ deconfined phase}\;.
% 		\left\{
%         \begin{array}{ll}
%             0 &\text{confined phase} \\
%             k ( 2\pi/\Nc) & \text{deconfined phase, $k = 0,\ldots, \Nc-1$}
%         \end{array}
%     \right.
% % \end{cases}
% \end{equation}

For $\MuI < \Temp \pi/\Nc$, the chiral and deconfinement critical lines $m_c(\Mu)$ emanate from the two tricritical points and continue to $\Mu \geq 0$, thus constraining the physical phase diagram \cite{deForcrand:2010he,Bonati:2010gi}.
Mapping the chiral critical line may also allow to clarify the nature of the transition in
the $\Nf=2$ chiral limit, see e.g. \cite{Bonati:2013tqa}. 

\section{Lattice action, observables and simulation parameters}
\label{ch:lattice}

For this study we employ the standard Wilson gauge action,
\begin{equation}
	\label{WilsonGaugeAction}	
	\SGauge = \LatCoupling \sum_\LatX  \sum_{\mu, \nu > \mu} \left\{ 1 - \real \TrC (\Plaq(\LatX) )\right\} \;,
\end{equation}
with plaquette \Plaq\ and lattice coupling $\LatCoupling = \frac{2 \Nc}{\GSq}$,
lattice sites \LatX\ and Dirac indices $\mu, \nu$.
We consider $\Nf=2$  mass-degenerate quarks with the standard Wilson 
action
\begin{equation}
	\label{SFermLatGen}
	\SFermion [\psibar, \psi, \Link ] = \LatSpacing^{4} \sum_{\Nf}\sum_{\LatX,\LatY} \psibar(\LatX)\ \MFermion(\LatX, \LatY) [\Link ]\ \psi(\LatY)\;,
\end{equation}
with fermion matrix
\begin{align}
	\label{MWilson}
	\MFermion(\LatX, \LatY) 	=&\  \delta_{\LatX\LatY}
	- \kappa \sum_{i=1}^{3} \left\{	(1 - \gamma_i)\Link_{\pm i}(\LatX)\delta_{\LatX+\hat{i},\LatY}	 \right\} \nonumber \\
					-&\ \kappa  \Bigg\{	(1 - \gamma_0)\ \ee^{+\LatSpacing\Mu}\Link_0(\LatX)\delta_{\LatY, \LatX+\hat{0}} \nonumber \\
				&\;\;\;\;\;+ 	(1 + \gamma_0)\ \ee^{-\LatSpacing\Mu}\LinkDag_0(\LatY)\delta_{\LatY, \LatX-\hat{0}}	 \Bigg\} \;.
\end{align}
Shorthand notation $\gamma_{-\mu} = - \gamma_\mu$ and $\Link_{-\mu}(\LatX) = \Link^{\dagger}_\mu(\LatX - \vec\mu)$ has been used.
In this formulation, the bare fermion mass \LatMass\ is encapsulated in the hopping parameter
\begin{equation}
	\label{LatMassWilson}
	\kappa = (2(\LatSpacing \LatMass + 4))^{-1} \;.
\end{equation}
Finite temperature on the lattice is given by 
\begin{equation}
	\label{LatTemp}
	\Temp = 1/\left( \LatSpacing(\LatCoupling)\NTau \right)\;.
\end{equation}
As observables we use the order parameters for center and chiral symmetry breaking, respectively, i.e.~the Polyakov loop at spatial 
site \LatXSpatial\ 
\begin{equation}
	\label{PolyakovLoopLattice}
	L (\LatXSpatial)= \frac{1}{\VSpatial}\TrC \Pi_{x_0 = 0}^{\NTau-1} \Link_0 (x_0,\LatXSpatial) \;,
\end{equation}
and the chiral condensate
\begin{equation}
	\label{pbp_1}
	\chiralcond = \Nf\ \Tr{\MFermionMinus}\;.
\end{equation}
Non-analytic phase transitions only exist
in the thermodynamic limit $\VSpatial \rightarrow \infty$.
To extract them from finite volume simulations, an extrapolation including
a finite size scaling study must be employed.
We use the Binder cumulant \cite{Binder:1981sa} constructed
from $\GenObs=\PolyIm$, 
\begin{equation}
	\Binder(\GenObs) = \langle (\GenObs - \langle \GenObs \rangle)^{4}\rangle /  \langle	(\GenObs - \langle \GenObs \rangle )^{2}	\rangle^{2}\;.
	\label{bindercum}
\end{equation}
Its value 
in the thermodynamic limit for different orders of the phase transition
is summarized in Table \ref{tab:bindervalues}.
The leading finite-size corrections are given by a Taylor expansion (cf. \cite{deForcrand:2010he})
\begin{align}
	\Binder(\LatCoupling, \NSigma) &= \Binder(\LatCoupling, \infty) + a_1 (\LatCoupling - \LatCoupling_c) \NSigma^{1/\nu} \nonumber \\
	& + a_2 ((\LatCoupling - \LatCoupling_c) \NSigma^{1/\nu})^{2} + \ldots\;.
	\label{BinderScaling}
\end{align}
Alternatively, the transition temperature may be extracted from the peak of the susceptibility
\begin{equation}
	\label{PolyAbsSusc}
	\Susc(\GenObs) = \VSpatial \langle ( \GenObs - \langle \GenObs \rangle)^2 \rangle\;.
\end{equation}
In the vicinity of the transition point, \Susc\ is expected to scale 
according (cf.~\cite{Bonati:2010gi}):
\begin{equation}
	\Susc = \NSigma^{\gamma/\nu} f(t\NSigma^{1/\nu})\;.
	\label{suscUniversalScaling}
\end{equation}
Here, $f$ is a universal scaling function, $t$ is the reduced temperature $t = (\Temp - \Tc)/\Tc$ and
$\gamma, \nu$ are critical exponents specific to the universality class of the transition \footnote{Strictly speaking, critical exponents can be defined for second-order transitions only. However, similar considerations can be carried out for the first-order case, too, cf. \cite{Fisher:1982xt}.}
 (see Table \ref{tab:bindervalues}).
With $f$ unknown, the critical exponents can be estimated by looking at 
$\Susc / \NSigma^{\gamma/\nu}$ against $t\NSigma^{1/\nu}$ for multiple spatial volumes.
These curves should coincide for the correct values of $\nu$ and $\gamma$ (collapse plot) and we use them to check the values for $\nu$ determined
from the Binder cumulant.

All simulations presented below were carried out using the
OpenCL \cite{opencl} based code \codename\ \cite{Bach:2012iw},
which runs efficiently on Graphic Processing Units (GPUs)
on 
\Loewe\ \cite{Bach2011a} at Goethe-University Frankfurt and on 
\Sanam\ at GSI Darmstadt (see e.g. \cite{sanam}).
We work at fixed temporal lattice extent $\NTau = 4$ and 
$\MuIc = \pi \Temp$. 
\begin{figure}
	\centering
	\includegraphics[width=\columnwidth]{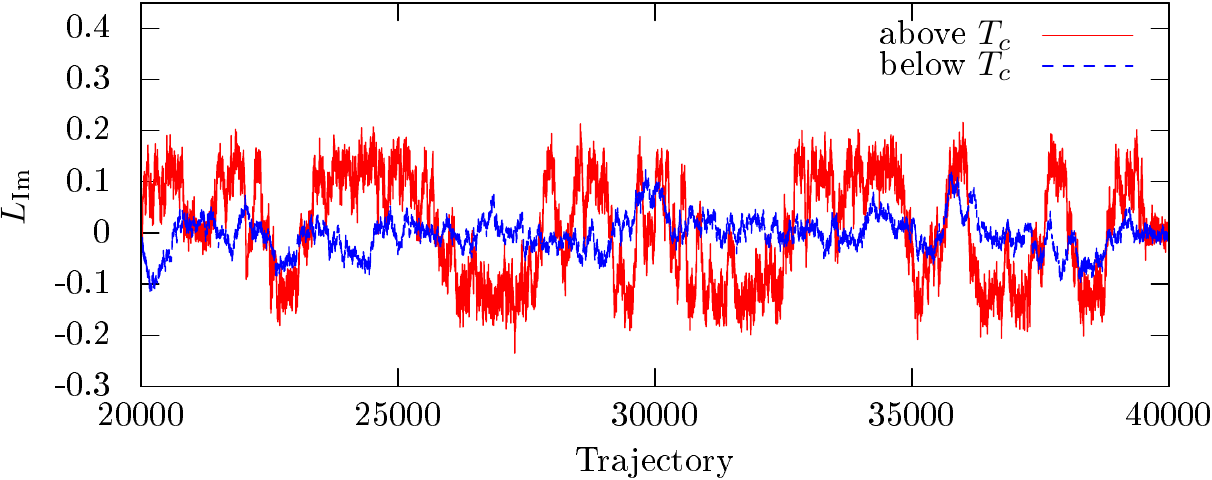}
	\caption{
					HMC history of \PolyIm\ at $\LatMassWilson = 0.1$ and $\NSigma = 12$, above and below \Tc 
					}
	\label{fig:polyImExample}
\end{figure}
In order to determine the phase diagram, Figure \ref{fig:rw_endpoint}, we
simulated 24 mass values ranging from $\LatMassWilson = 0.03 \ldots 0.165$.
For the finite size scaling, each \LatMassWilson\ was simulated on at least three, in some cases four or five spatial volumes,
ranging from $\NSigma=8$ to $20$.
To scan the temperature,
at least ten \LatCoupling-values with $\Delta\LatCoupling = 0.001$ around 
\Tc\ have been simulated on each lattice.
In each run, 35k HMC trajectories of unit length have been 
produced after 5k trajectories of thermalization. 
In some cases this number has been extended to 75k.
The acceptance rate in each run was of the order of $75\%$.
Additional \LatCoupling-points have been filled in using Ferrenberg-Swendsen reweighting \cite{Ferrenberg:1989ui}.
Details about the simulations can be found in Table \ref{tab:muiSims} in the \appendixname.
\begin{table}[ht]
	\centering
%this produces a stucked float, but I think its alright...
\begin{tabular}{|c|cccc|}
\hline 
& Crossover & 3D Ising & triple point & tricritical \\
\hline
$\Binder(X)$ & 3 & 1.604 & 1.5 & 2 \\
\hline
$\nu$ & - & 0.6301(4) & 1/3 & 1/2 \\
\hline
$\gamma$ & - & 1.2372(5) & 1 & 1 \\
\hline
\end{tabular} 
\caption{
	Values for the Binder cumulant $\Binder(X)$ \cite{deForcrand:2010he} and critical exponent $\nu$ for different phase transitions \cite{Pelissetto:2000ek}.
	}
% \vspace*{-.8cm}
\label{tab:bindervalues}
\end{table}

\section{Numerical results}
\label{ch:results_wilson}

\begin{figure}
 		\centering
   \includegraphics[width=\columnwidth]{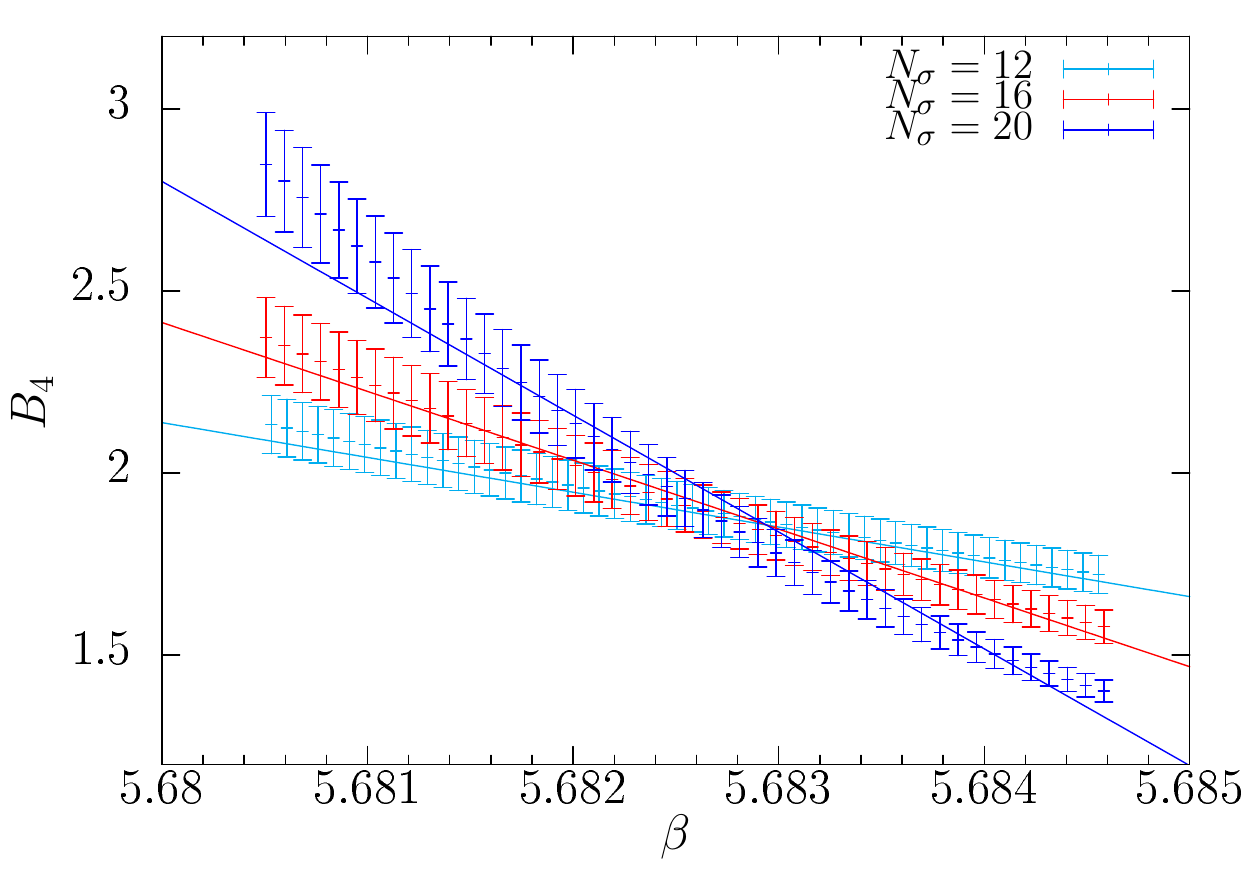}
 		\caption{Reweighted Binder Cumulant of \PolyIm\ at $\LatMassWilson = 0.07$ for various \NSigma, including the fits to the finite size scaling form.}
 		\label{fig:scalingBinder}
\end{figure}
\begin{figure}
	\centering
	\includegraphics[width=\columnwidth]{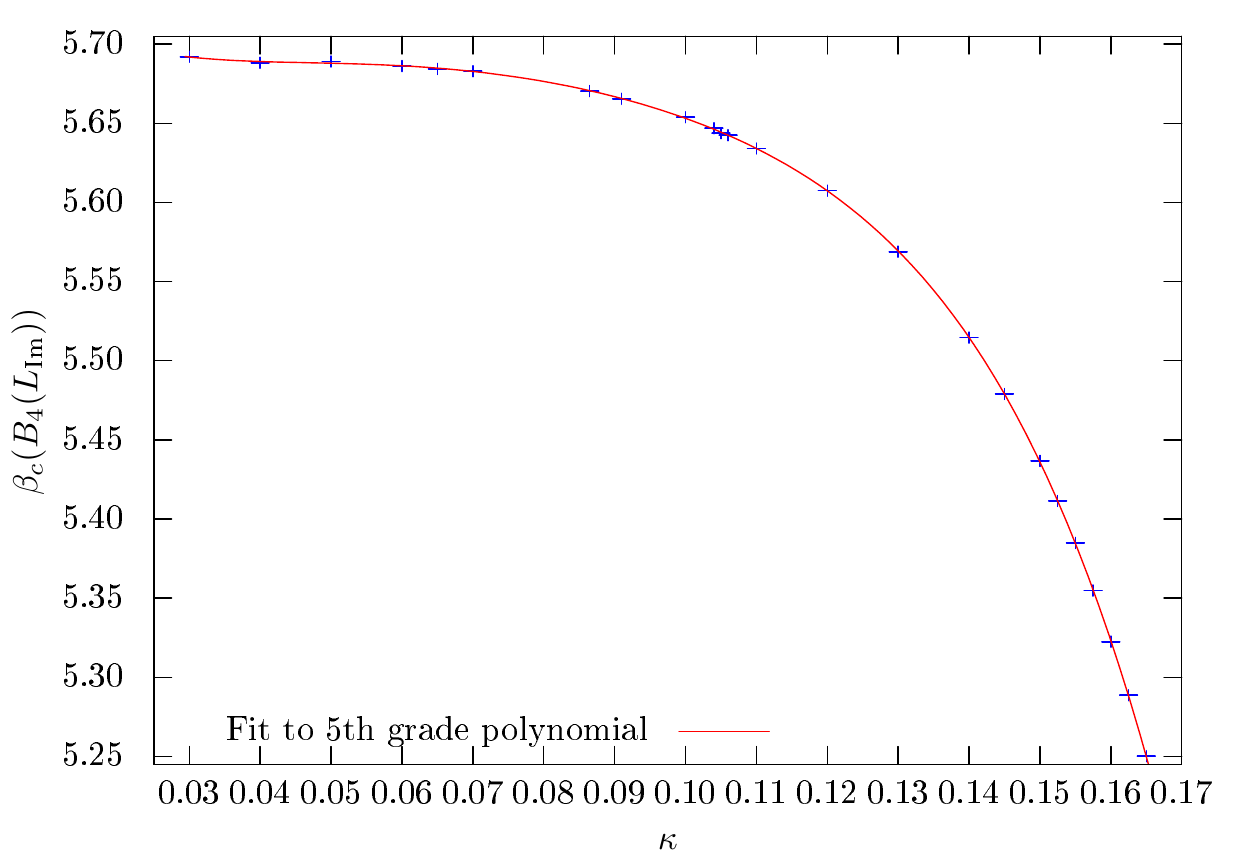}
	\caption{
					\LatCouplingC\ as a function of \LatMassWilson, extracted from fits to \Binder\ data according to \eqref{BinderScaling}.	
					Also shown is a fit to a fifth 
grade polynomial.
					}
	\label{fig:fittedBetaC}
\end{figure}
\begin{figure}
	\centering
  \includegraphics[width=\columnwidth]{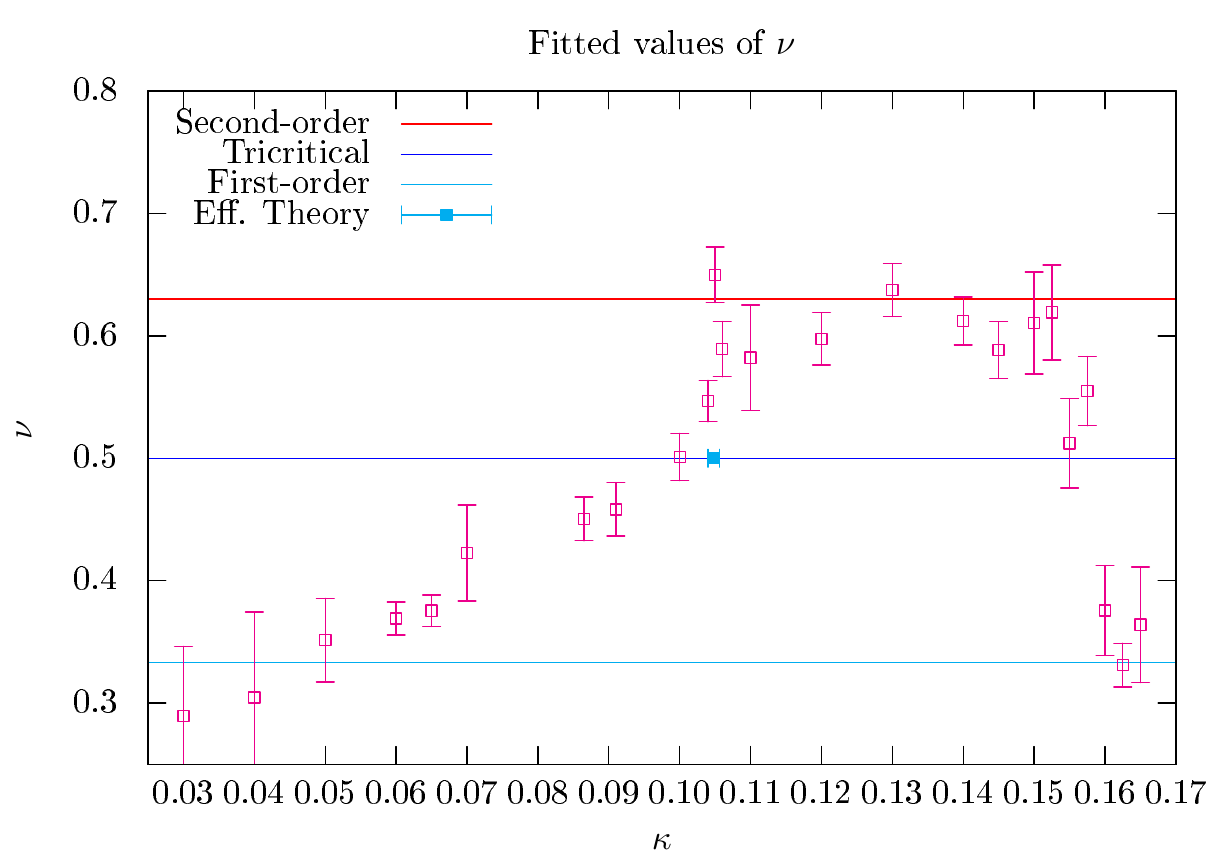}
	\caption{
					Fitted critical exponent $\nu$ as a function of \LatMassWilson.
					Also shown are values of $\nu$ for certain universality classes as well as the prediction for the tricritical mass from the effective theory \cite{Fromm:2011qi}.
					}
	\label{fig:KappaNu}
\end{figure}

At fixed value of \MuI\  on the boundary between center sectors, 
the phase of the Polyakov loop, or its imaginary part \PolyIm, fluctuates between the values realized in each sector.
At high temperatures these fluctuations are jumps between two distinct non-zero
values, while for lower temperatures they are realized smoothly around zero, 
as shown in Figure \ref{fig:polyImExample}.
Note that in both cases one has $\PolyImExp=0$ for sufficiently large 
statistics. For high temperatures (in the first-order case), 
the system will eventually take 
on one of the possible values in the thermodynamic limit, i.e.~the symmetry
will break spontaneously.
In the following, we study the nature of the RW endpoint as a function of 
the fermion mass 
in analogy to the staggered study \cite{deForcrand:2010he}.

Figure \ref{fig:scalingBinder} shows the functional behavior of the Binder 
cumulant \Binder\ 
for one particular quark mass as the spatial volume is increased.
\Binder\ decreases with \LatCoupling\ and gets steeper as the volume is increased.
This is expected as below and above \LatCouplingC\ 
a crossover and first-order region is located, which have a \Binder\ value 
of 3 and 1.5, respectively, in the thermodynamic limit, where
 \Binder\ approaches a step function.
The intersection of the three finite volumes gives an estimate for the location of the RW endpoint. 
To extract it together with the critical exponent $\nu$, we fit
to the scaling form \eqref{BinderScaling}.
The resulting value for $\Binder(\LatCoupling, \infty)$ is found to be 
somewhat higher than the universal values because of large 
finite volume corrections, in agreement with the observations in staggered
simulations \cite{deForcrand:2010he}.
The critical exponent $\nu$, however, can be extracted quite well.
This procedure is carried out for all simulated values of \LatMassWilson\ 
and the results for the critical coupling and exponent are 
collected in Figures \ref{fig:fittedBetaC} and \ref{fig:KappaNu}, respectively.

\begin{figure*}
        \centering
 \subfigure{
    \includegraphics[width=.95\columnwidth]{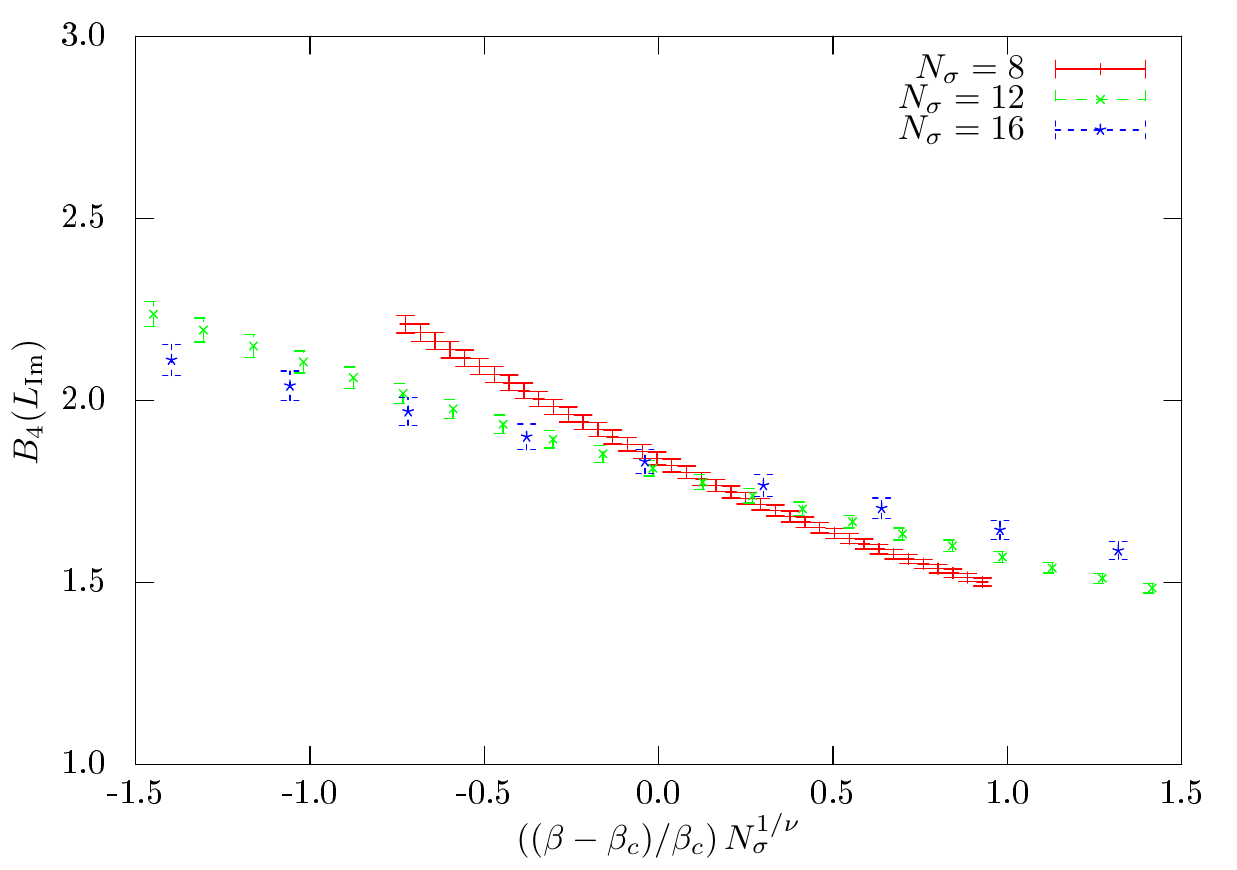}
    \includegraphics[width=.95\columnwidth]{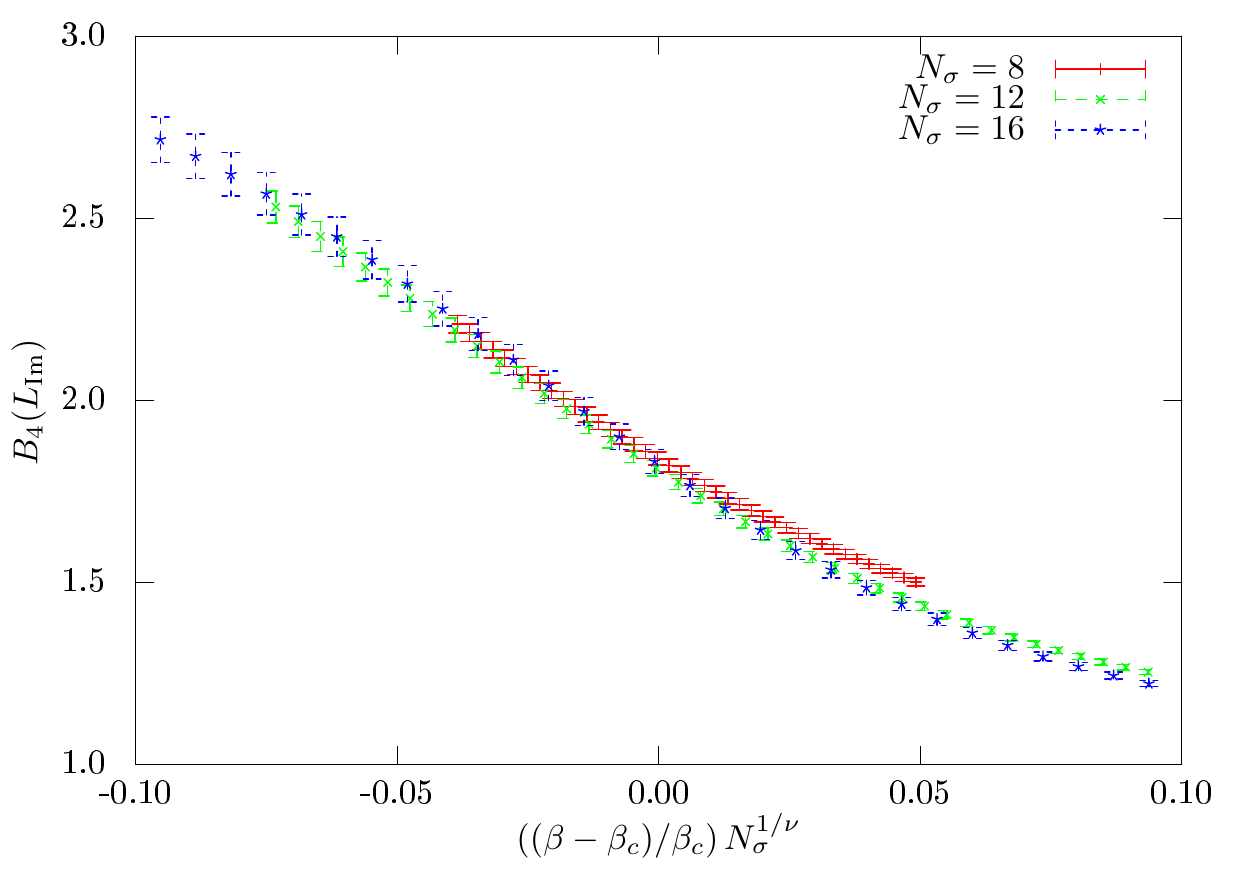}
	}
	\caption{
     Collapse plots of $\Binder(\PolyIm)$ at  at \LatMassWilson\ = 0.130 
for first- (left) and second-order (right) critical exponents.
   }
 \label{fig:k1300_binder_collapse}
\end{figure*}

\begin{figure*}
        \centering
 \subfigure{
    \includegraphics[width=.95\columnwidth]{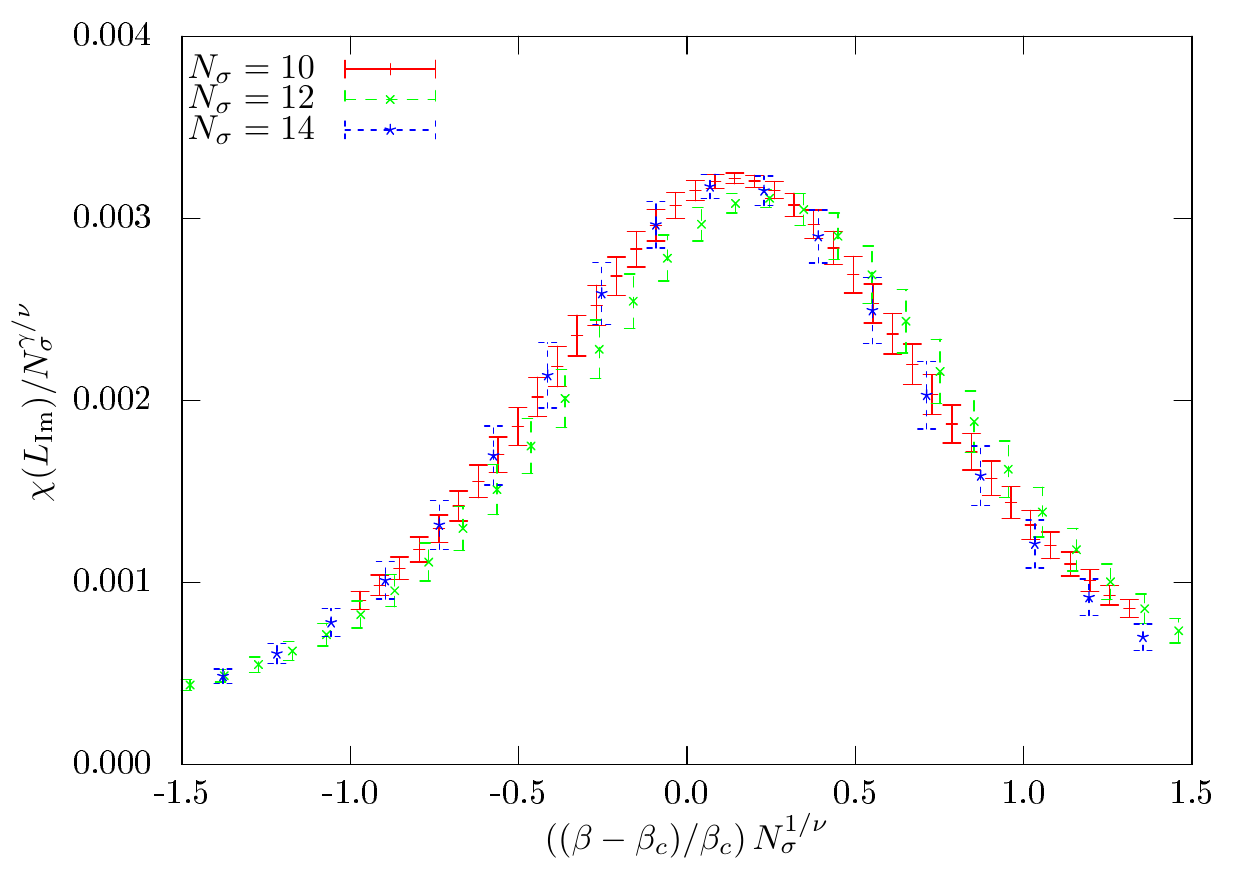}
    \includegraphics[width=.95\columnwidth]{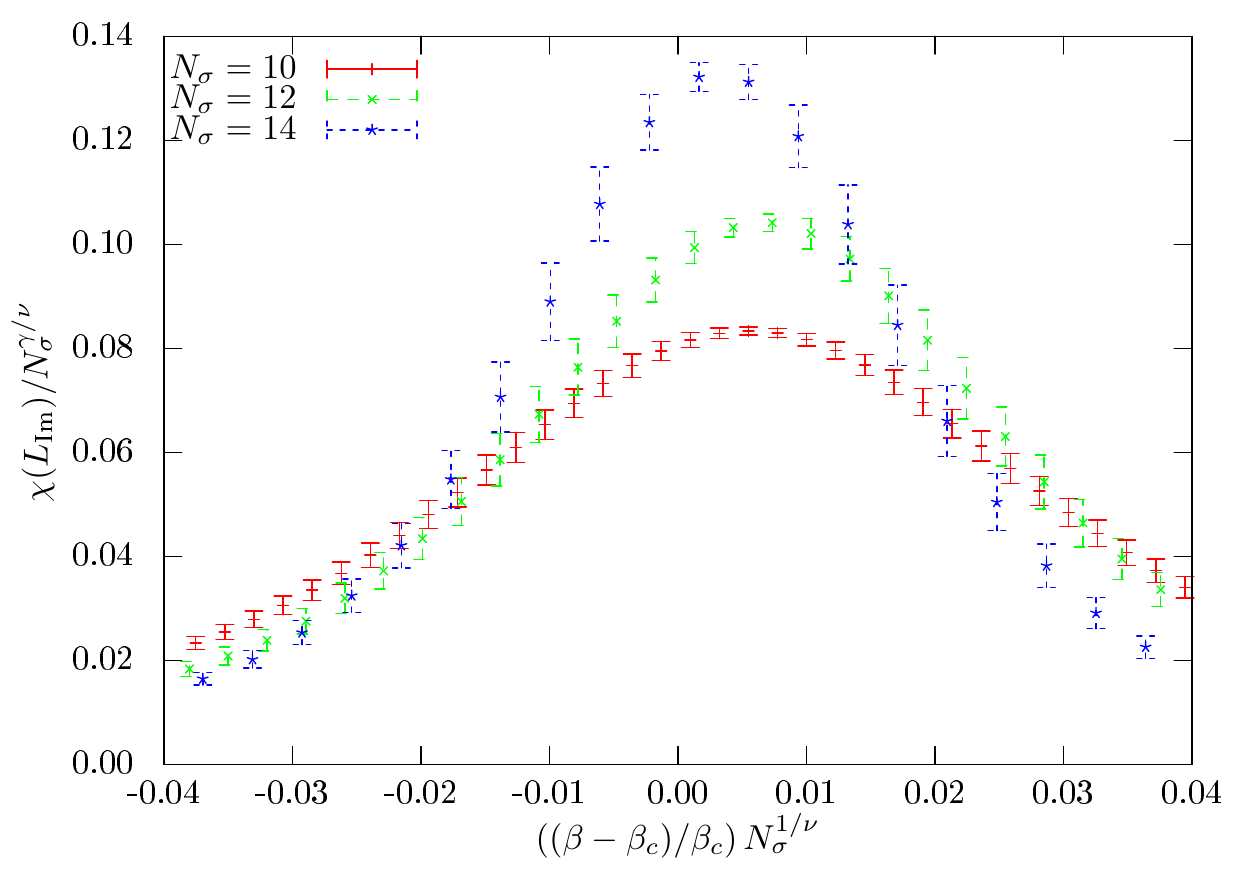}
	}
	\caption{
     Collapse plots of $\Susc(\PolyIm)$  at \LatMassWilson\ = 0.165 for 
first- (left) and second-order (right) critical exponents.
   }
 \label{fig:k1650_susc_collapse}
\end{figure*}

\begin{figure*}
        \centering
 \subfigure{
    \includegraphics[width=.95\columnwidth]{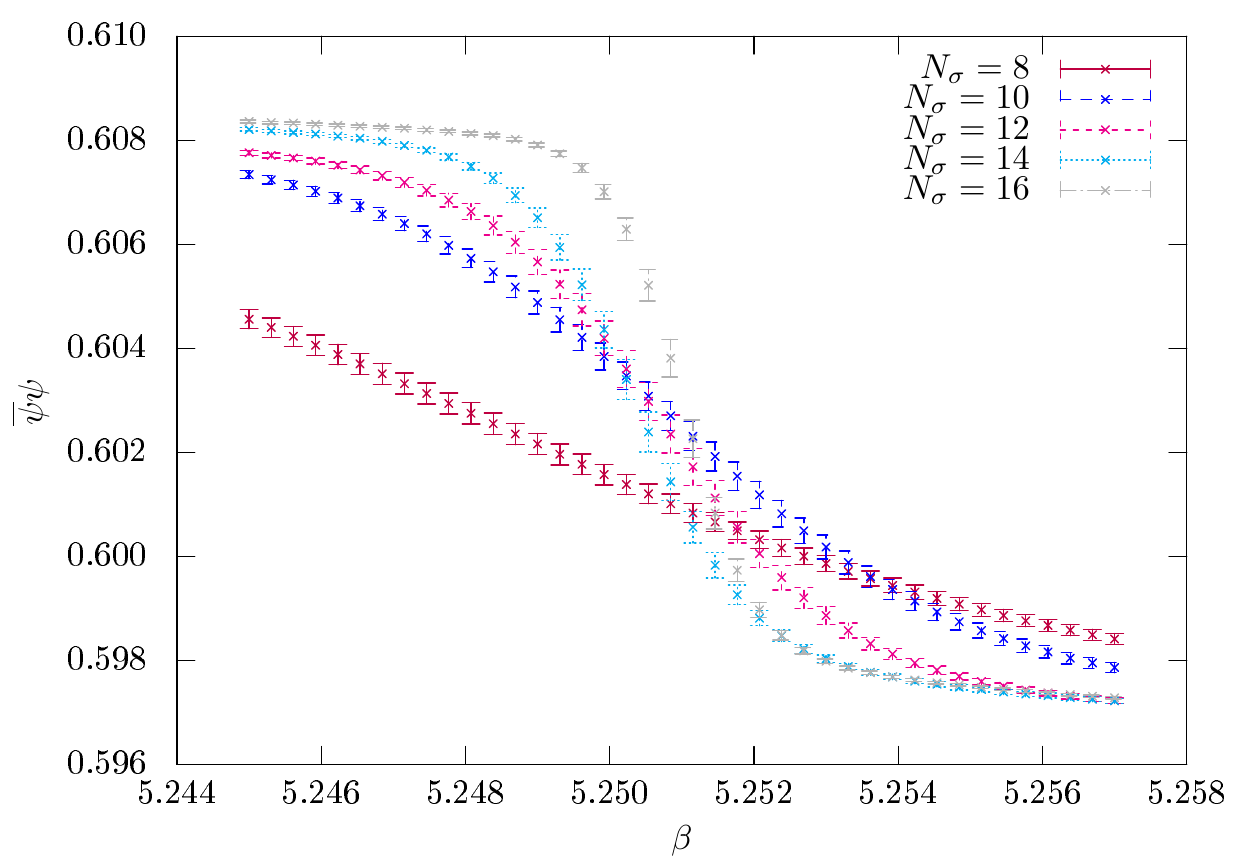}
		\includegraphics[width=.95\columnwidth]{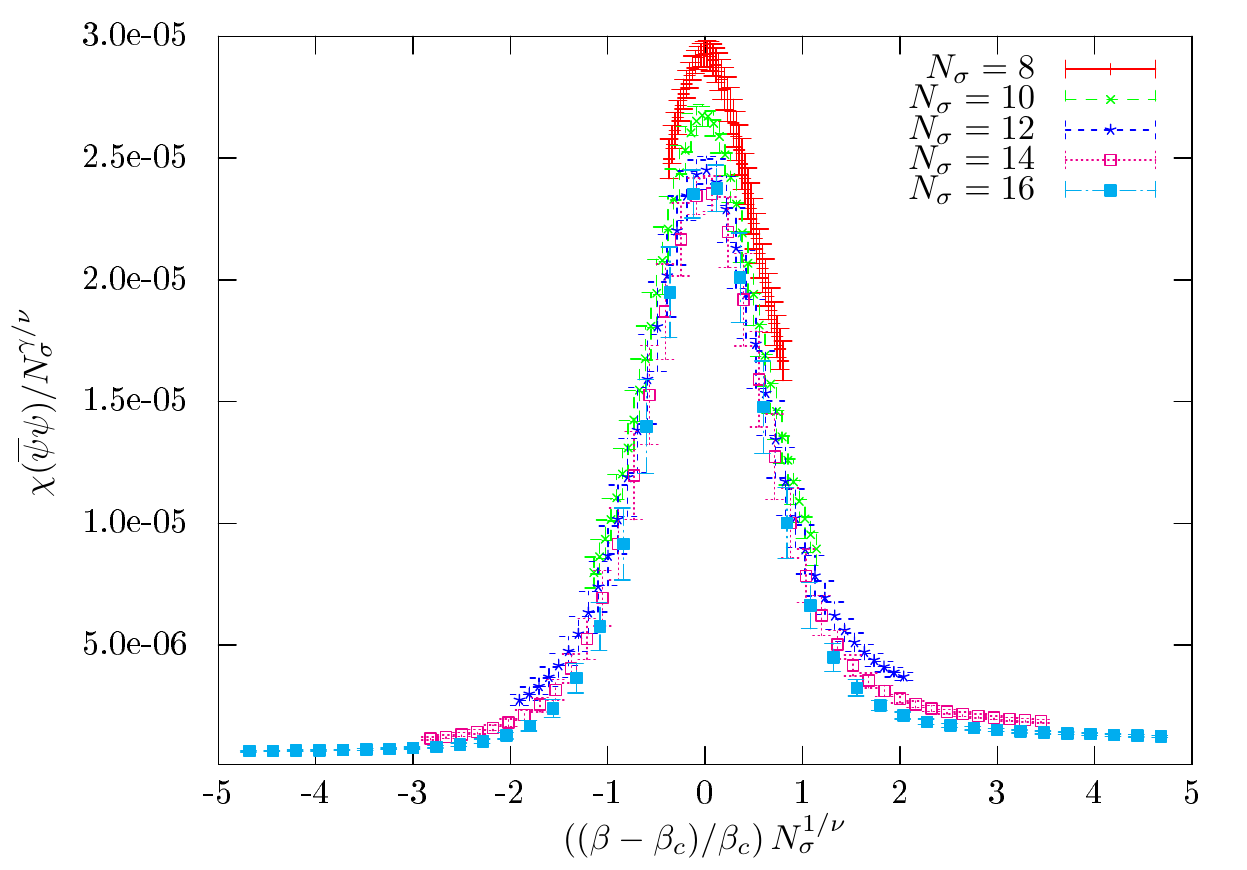}
	}
	\caption{
     \chiralcond\ at $\LatMassWilson = 0.165$ as \NSigma\ is increased (left). 
			Collapse plot of $\Susc(\chiralcond)$ at $\LatMassWilson = 0.165$ according to first-order exponents (right).
   }
 \label{fig:pbpScaling}
\end{figure*}

Note how \LatCouplingC\ shows a significant fermionic influence for 
$\LatMassWilson \gtrsim 0.085$ only, in accord with the nature of
the phase transition, which stays first-order as in pure gauge theory
in the large quark mass range.
For intermediate masses, the RW endpoint is of second-order and 
changes back to triple point nature in the light mass region, 
where the RW transition meets with a first-order chiral transition.

The analysis of the Binder cumulant can be checked and confirmed
by also looking at the susceptibilities of various observables.
For \PolyAbs, \PolyIm\ and \chiralcond\, 
fully consistent values for \LatCouplingC\ 
are found from the peak of the susceptibility, see Tables \ref{tab:muiSims_betaCDeconf},\ \ref{tab:muiSims_betaCPolyIm} and \ref{tab:muiSims_betaCPbp} in the \appendixname.

The identification of the order of the transition can also be checked by collapse
plots using the scaling form of \Binder\ as well as the
susceptibilities. 
Figures \ref{fig:k1300_binder_collapse} and \ref{fig:k1650_susc_collapse} show 
examples 
for $\LatMassWilson = 0.130$ and $0.165$, respectively, allowing for a clear
discrimination between different scaling scenarios,
with $\LatMassWilson = 0.130$ in the second-order region and 
$\LatMassWilson = 0.165$ in the first-order region.

For lighter masses,
the chiral transition is also seen in the melting of the chiral condensate \chiralcond\ and the peak of its susceptibility, respectively.
This is shown in Figure \ref{fig:pbpScaling} (left), which shows \chiralcond\ at $\LatMassWilson = 0.165$ for different \NSigma.
As the volume of the system is increased, the gradient gets steeper, as expected to happen in a first-order transition. 
In addition, the collapse plot of $\Susc(\chiralcond)$ in Figure \ref{fig:pbpScaling} (right)  
clearly confirms first-order behavior.
Our results in the small mass region partly support earlier ones from 
\cite{Wu:2013bfa}, where
the authors simulated at various $\LatMassWilson \geq 0.155$ and 
find that these all lie in the first-order region. However, no
analysis for a tricritical point has been carried out there.

Note that determining the order of the transition from 
collapse plots can be inconclusive if the exponents for different scenarios 
take on similar values, as is the case for second-order and 
tricritical exponents: $\gamma/\nu = $ 1.963 and 2, respectively. 
The identification of the tricritical points separating the triple point 
regions from second-order endpoint regions is thus best carried out using 
the Binder cumulant, cf. Figure \ref{fig:KappaNu}.

We thus determine the two tricritical \LatMassWilson\ to be:
\begin{eqnarray}
	\label{tricriticalKappa}
	&\LatMassWilsonTricHeavy &= 0.1000 \pm 0.090\;,\\
	\label{tricriticalKappaHeavy}
	&\LatMassWilsonTricLight &= 0.1550 \pm 0.050\;.
	\label{tricriticalKappaLight}
\end{eqnarray}
The errors on these values are chosen conservatively such that 
the neighboring simulation points which clearly fall into the first and 
second-order scenario, respectively, are taken as a boundary 
for the tricritical masses. In summary, the qualitative phase structure for 
the RW transition is exactly as for staggered 
fermions \cite{deForcrand:2010he,Bonati:2010gi}.

Our results for the location of the tricritical points on $N_\tau=4$ may also
serve as a quantitative check for the predictions of an effective lattice
theory for finite density \cite{Fromm:2011qi}, which includes the 
fermion determinant through order $\kappa^2$ only. There, the 
tricritical point in the heavy mass region was predicted to be
\begin{equation}
	\LatMassWilsonTricHeavy (\text{eff. theory}) = 0.1048 \pm 0.0008\;,
	\label{effModelPrediction}
\end{equation}
in good agreement with the full \LQcd\ simulations.

Finally, we
estimate the pion mass \mpi\  at the light tricritical point. To this end, $\Temp,\Mu=0$ simulations at
 the tricritical couplings $\beta=5.355, \LatMassWilson = 0.1575$ were run on a $16^3\times32$ lattice on 
 \Juqueen\footnote{http://www.fz-juelich.de/ias/jsc/EN/Expertise/\\Supercomputers/JUQUEEN/JUQUEEN\_node.html}.
We generated 4200 trajectories after 500 trajectories of thermalization.
On these, the effective masses for the pion and rho particle were estimated to 
be $\LatSpacing\mpi=1.1426(17)$ and 
$\LatSpacing m_\rho=1.2147(25)$\footnote{We thank Georg Bergner for help with this analysis.}. 
Equating $\LatSpacing m_\rho$ to the physical value yields $\mpi \approx 729(2)$ MeV and a lattice spacing of $\LatSpacing \approx 0.3$ fm.
This corresponds to a ratio $\mpi/m_\rho = 0.94064$,
compared to the physical value of $0.18003$, i.e.~on coarse lattices the
chiral first-order RW region is very wide.

\section{Summary and Perspectives}

We have performed a study of $N_f=2$ QCD
at imaginary value $\mu=i\pi T/3$ 
of the quark chemical potential, confirming an 
interesting phase structure seen earlier in simulations using staggered 
fermions.
In particular, it was found that the endpoint of the transition between
different center sectors is connected to the analytic continuation of
the deconfinement and chiral transitions, and hence its nature 
depends non-trivially 
on the quark mass as in Figure \ref{fig:rw_endpoint}. First-order
regions at small and large quark masses are 
separated by two tricritical points
from a second-order region at intermediate quark masses.
In addition, we observe good agreement between the prediction of an effective lattice theory for the 
large mass tricritical point and our full simulation result.
Cut-off effects on the location of the tricritical points
are expected to be strong and can be studied by increasing \NTau, at
significantly larger numerical cost. It might also be interesting 
to follow \cite{Bonati:2013tqa} and repeat the simulations at smaller 
values of imaginary chemical potential in order
to determine the nature of the $\mu=0$ transition in the chiral limit. 

% \begin{figure}
% 	\centering
% 	\includegraphics[width=\columnwidth]{./pbp_mean_rew}
% 	\caption{
% 					\chiralcond\ at $\LatMassWilson = 0.165$ as \NSigma\ is increased.
% 					}
% 	\label{fig:pbpScaling}
% \end{figure}
% 
% \begin{figure}
% 	\centering
% 	\includegraphics[width=\columnwidth]{./pbp_susc_rew_collapse1o}
% 	\caption{
% 					Collapse plot of $\Susc(\chiralcond)$ at $\LatMassWilson = 0.165$ according to first-order exponents.
% 					}
% 	\label{fig:pbpScalingSusc}
% \end{figure}
%
%

\acknowledgments

O. P, and C. P. are supported by the Helmholtz International Center for FAIR within the LOEWE program of the State of Hesse.
C.P. is supported by the GSI Helmholtzzentrum f\"{u}r Schwerionenforschung.

%%%%%%%%%%%%%%%%%%%%%%%%%%%%%%%%%%%%%%%%
% bibliography
\bibliographystyle{apsrev4-1}
\bibliography{./RW_nf2_wilson.bib}

\begin{thebibliography}{28}%
\makeatletter
\providecommand \@ifxundefined [1]{%
 \@ifx{#1\undefined}
}%
\providecommand \@ifnum [1]{%
 \ifnum #1\expandafter \@firstoftwo
 \else \expandafter \@secondoftwo
 \fi
}%
\providecommand \@ifx [1]{%
 \ifx #1\expandafter \@firstoftwo
 \else \expandafter \@secondoftwo
 \fi
}%
\providecommand \natexlab [1]{#1}%
\providecommand \enquote  [1]{``#1''}%
\providecommand \bibnamefont  [1]{#1}%
\providecommand \bibfnamefont [1]{#1}%
\providecommand \citenamefont [1]{#1}%
\providecommand \href@noop [0]{\@secondoftwo}%
\providecommand \href [0]{\begingroup \@sanitize@url \@href}%
\providecommand \@href[1]{\@@startlink{#1}\@@href}%
\providecommand \@@href[1]{\endgroup#1\@@endlink}%
\providecommand \@sanitize@url [0]{\catcode `\\12\catcode `\$12\catcode
  `\&12\catcode `\#12\catcode `\^12\catcode `\_12\catcode `\%12\relax}%
\providecommand \@@startlink[1]{}%
\providecommand \@@endlink[0]{}%
\providecommand \url  [0]{\begingroup\@sanitize@url \@url }%
\providecommand \@url [1]{\endgroup\@href {#1}{\urlprefix }}%
\providecommand \urlprefix  [0]{URL }%
\providecommand \Eprint [0]{\href }%
\providecommand \doibase [0]{http://dx.doi.org/}%
\providecommand \selectlanguage [0]{\@gobble}%
\providecommand \bibinfo  [0]{\@secondoftwo}%
\providecommand \bibfield  [0]{\@secondoftwo}%
\providecommand \translation [1]{[#1]}%
\providecommand \BibitemOpen [0]{}%
\providecommand \bibitemStop [0]{}%
\providecommand \bibitemNoStop [0]{.\EOS\space}%
\providecommand \EOS [0]{\spacefactor3000\relax}%
\providecommand \BibitemShut  [1]{\csname bibitem#1\endcsname}%
\let\auto@bib@innerbib\@empty
%</preamble>
\bibitem [{\citenamefont {Aoki}\ \emph {et~al.}(2006)\citenamefont {Aoki},
  \citenamefont {Endrodi}, \citenamefont {Fodor}, \citenamefont {Katz},\ and\
  \citenamefont {Szabo}}]{Aoki:2006we}%
  \BibitemOpen
  \bibfield  {author} {\bibinfo {author} {\bibfnamefont {Y.}~\bibnamefont
  {Aoki}}, \bibinfo {author} {\bibfnamefont {G.}~\bibnamefont {Endrodi}},
  \bibinfo {author} {\bibfnamefont {Z.}~\bibnamefont {Fodor}}, \bibinfo
  {author} {\bibfnamefont {S.}~\bibnamefont {Katz}}, \ and\ \bibinfo {author}
  {\bibfnamefont {K.}~\bibnamefont {Szabo}},\ }\href {\doibase
  10.1038/nature05120} {\bibfield  {journal} {\bibinfo  {journal} {Nature}\
  }\textbf {\bibinfo {volume} {443}},\ \bibinfo {pages} {675} (\bibinfo {year}
  {2006})},\ \Eprint {http://arxiv.org/abs/hep-lat/0611014}
  {arXiv:hep-lat/0611014 [hep-lat]} \BibitemShut {NoStop}%
%%CITATION = HEP-LAT/0611014;%%
\bibitem [{\citenamefont {de~Forcrand}\ and\ \citenamefont
  {Philipsen}(2007)}]{deForcrand:2006pv}%
  \BibitemOpen
  \bibfield  {author} {\bibinfo {author} {\bibfnamefont {P.}~\bibnamefont
  {de~Forcrand}}\ and\ \bibinfo {author} {\bibfnamefont {O.}~\bibnamefont
  {Philipsen}},\ }\href {\doibase 10.1088/1126-6708/2007/01/077} {\bibfield
  {journal} {\bibinfo  {journal} {JHEP}\ }\textbf {\bibinfo {volume} {0701}},\
  \bibinfo {pages} {077} (\bibinfo {year} {2007})},\ \Eprint
  {http://arxiv.org/abs/hep-lat/0607017} {arXiv:hep-lat/0607017 [hep-lat]}
  \BibitemShut {NoStop}%
%%CITATION = HEP-LAT/0607017;%%
\bibitem [{\citenamefont {Philipsen}(2012)}]{Philipsen:2011zx}%
  \BibitemOpen
  \bibfield  {author} {\bibinfo {author} {\bibfnamefont {O.}~\bibnamefont
  {Philipsen}},\ }\href {\doibase 10.5506/APhysPolBSupp.5.825} {\bibfield
  {journal} {\bibinfo  {journal} {Acta Phys.Polon.Supp.}\ }\textbf {\bibinfo
  {volume} {5}},\ \bibinfo {pages} {825} (\bibinfo {year} {2012})},\ \Eprint
  {http://arxiv.org/abs/1111.5370} {arXiv:1111.5370 [hep-ph]} \BibitemShut
  {NoStop}%
%%CITATION = ARXIV:1111.5370;%%
\bibitem [{\citenamefont {de~Forcrand}\ and\ \citenamefont
  {Philipsen}(2010)}]{deForcrand:2010he}%
  \BibitemOpen
  \bibfield  {author} {\bibinfo {author} {\bibfnamefont {P.}~\bibnamefont
  {de~Forcrand}}\ and\ \bibinfo {author} {\bibfnamefont {O.}~\bibnamefont
  {Philipsen}},\ }\href {\doibase 10.1103/PhysRevLett.105.152001} {\bibfield
  {journal} {\bibinfo  {journal} {Phys.Rev.Lett.}\ }\textbf {\bibinfo {volume}
  {105}},\ \bibinfo {pages} {152001} (\bibinfo {year} {2010})},\ \Eprint
  {http://arxiv.org/abs/1004.3144} {arXiv:1004.3144 [hep-lat]} \BibitemShut
  {NoStop}%
%%CITATION = ARXIV:1004.3144;%%
\bibitem [{\citenamefont {Bonati}\ \emph {et~al.}(2011)\citenamefont {Bonati},
  \citenamefont {Cossu}, \citenamefont {D'Elia},\ and\ \citenamefont
  {Sanfilippo}}]{Bonati:2010gi}%
  \BibitemOpen
  \bibfield  {author} {\bibinfo {author} {\bibfnamefont {C.}~\bibnamefont
  {Bonati}}, \bibinfo {author} {\bibfnamefont {G.}~\bibnamefont {Cossu}},
  \bibinfo {author} {\bibfnamefont {M.}~\bibnamefont {D'Elia}}, \ and\ \bibinfo
  {author} {\bibfnamefont {F.}~\bibnamefont {Sanfilippo}},\ }\href {\doibase
  10.1103/PhysRevD.83.054505} {\bibfield  {journal} {\bibinfo  {journal}
  {Phys.Rev.}\ }\textbf {\bibinfo {volume} {D83}},\ \bibinfo {pages} {054505}
  (\bibinfo {year} {2011})},\ \Eprint {http://arxiv.org/abs/1011.4515}
  {arXiv:1011.4515 [hep-lat]} \BibitemShut {NoStop}%
%%CITATION = ARXIV:1011.4515;%%
\bibitem [{\citenamefont {de~Forcrand}\ and\ \citenamefont
  {Philipsen}(2008)}]{deForcrand:2008vr}%
  \BibitemOpen
  \bibfield  {author} {\bibinfo {author} {\bibfnamefont {P.}~\bibnamefont
  {de~Forcrand}}\ and\ \bibinfo {author} {\bibfnamefont {O.}~\bibnamefont
  {Philipsen}},\ }\href {\doibase 10.1088/1126-6708/2008/11/012} {\bibfield
  {journal} {\bibinfo  {journal} {JHEP}\ }\textbf {\bibinfo {volume} {0811}},\
  \bibinfo {pages} {012} (\bibinfo {year} {2008})},\ \Eprint
  {http://arxiv.org/abs/0808.1096} {arXiv:0808.1096 [hep-lat]} \BibitemShut
  {NoStop}%
%%CITATION = ARXIV:0808.1096;%%
\bibitem [{\citenamefont {Nagata}\ and\ \citenamefont
  {Nakamura}(2011)}]{Nagata:2011yf}%
  \BibitemOpen
  \bibfield  {author} {\bibinfo {author} {\bibfnamefont {K.}~\bibnamefont
  {Nagata}}\ and\ \bibinfo {author} {\bibfnamefont {A.}~\bibnamefont
  {Nakamura}},\ }\href {\doibase 10.1103/PhysRevD.83.114507} {\bibfield
  {journal} {\bibinfo  {journal} {Phys.Rev.}\ }\textbf {\bibinfo {volume}
  {D83}},\ \bibinfo {pages} {114507} (\bibinfo {year} {2011})},\ \Eprint
  {http://arxiv.org/abs/1104.2142} {arXiv:1104.2142 [hep-lat]} \BibitemShut
  {NoStop}%
%%CITATION = ARXIV:1104.2142;%%
\bibitem [{\citenamefont {Nagata}\ and\ \citenamefont
  {Nakamura}(2012)}]{Nagata:2012pc}%
  \BibitemOpen
  \bibfield  {author} {\bibinfo {author} {\bibfnamefont {K.}~\bibnamefont
  {Nagata}}\ and\ \bibinfo {author} {\bibfnamefont {A.}~\bibnamefont
  {Nakamura}},\ }\href {\doibase 10.1007/JHEP04(2012)092} {\bibfield  {journal}
  {\bibinfo  {journal} {JHEP}\ }\textbf {\bibinfo {volume} {1204}},\ \bibinfo
  {pages} {092} (\bibinfo {year} {2012})},\ \Eprint
  {http://arxiv.org/abs/1201.2765} {arXiv:1201.2765 [hep-lat]} \BibitemShut
  {NoStop}%
%%CITATION = ARXIV:1201.2765;%%
\bibitem [{\citenamefont {Wu}\ and\ \citenamefont {Meng}(2013)}]{Wu:2013bfa}%
  \BibitemOpen
  \bibfield  {author} {\bibinfo {author} {\bibfnamefont {L.-K.}\ \bibnamefont
  {Wu}}\ and\ \bibinfo {author} {\bibfnamefont {X.-F.}\ \bibnamefont {Meng}},\
  }\href {\doibase 10.1103/PhysRevD.87.094508} {\bibfield  {journal} {\bibinfo
  {journal} {Phys.Rev.}\ }\textbf {\bibinfo {volume} {D87}},\ \bibinfo {pages}
  {094508} (\bibinfo {year} {2013})},\ \Eprint {http://arxiv.org/abs/1303.0336}
  {arXiv:1303.0336 [hep-lat]} \BibitemShut {NoStop}%
%%CITATION = ARXIV:1303.0336;%%
\bibitem [{\citenamefont {Alexandru}\ and\ \citenamefont
  {Li}(2013)}]{Alexandru:2013uaa}%
  \BibitemOpen
  \bibfield  {author} {\bibinfo {author} {\bibfnamefont {A.}~\bibnamefont
  {Alexandru}}\ and\ \bibinfo {author} {\bibfnamefont {A.}~\bibnamefont {Li}},\
  }\href@noop {} {\bibfield  {journal} {\bibinfo  {journal} {PoS}\ }\textbf
  {\bibinfo {volume} {LATTICE2013}},\ \bibinfo {pages} {208} (\bibinfo {year}
  {2013})},\ \Eprint {http://arxiv.org/abs/1312.1201} {arXiv:1312.1201
  [hep-lat]} \BibitemShut {NoStop}%
%%CITATION = ARXIV:1312.1201;%%
\bibitem [{\citenamefont {Sharpe}(2006)}]{Sharpe:2006re}%
  \BibitemOpen
  \bibfield  {author} {\bibinfo {author} {\bibfnamefont {S.~R.}\ \bibnamefont
  {Sharpe}},\ }\href@noop {} {\bibfield  {journal} {\bibinfo  {journal} {PoS}\
  }\textbf {\bibinfo {volume} {LAT2006}},\ \bibinfo {pages} {022} (\bibinfo
  {year} {2006})},\ \Eprint {http://arxiv.org/abs/hep-lat/0610094}
  {arXiv:hep-lat/0610094 [hep-lat]} \BibitemShut {NoStop}%
%%CITATION = HEP-LAT/0610094;%%
\bibitem [{\citenamefont {Fromm}\ \emph {et~al.}(2012)\citenamefont {Fromm},
  \citenamefont {Langelage}, \citenamefont {Lottini},\ and\ \citenamefont
  {Philipsen}}]{Fromm:2011qi}%
  \BibitemOpen
  \bibfield  {author} {\bibinfo {author} {\bibfnamefont {M.}~\bibnamefont
  {Fromm}}, \bibinfo {author} {\bibfnamefont {J.}~\bibnamefont {Langelage}},
  \bibinfo {author} {\bibfnamefont {S.}~\bibnamefont {Lottini}}, \ and\
  \bibinfo {author} {\bibfnamefont {O.}~\bibnamefont {Philipsen}},\ }\href
  {\doibase 10.1007/JHEP01(2012)042} {\bibfield  {journal} {\bibinfo  {journal}
  {JHEP}\ }\textbf {\bibinfo {volume} {1201}},\ \bibinfo {pages} {042}
  (\bibinfo {year} {2012})},\ \Eprint {http://arxiv.org/abs/1111.4953}
  {arXiv:1111.4953 [hep-lat]} \BibitemShut {NoStop}%
%%CITATION = ARXIV:1111.4953;%%
\bibitem [{\citenamefont {Roberge}\ and\ \citenamefont
  {Weiss}(1986)}]{Roberge:1986mm}%
  \BibitemOpen
  \bibfield  {author} {\bibinfo {author} {\bibfnamefont {A.}~\bibnamefont
  {Roberge}}\ and\ \bibinfo {author} {\bibfnamefont {N.}~\bibnamefont
  {Weiss}},\ }\href {\doibase 10.1016/0550-3213(86)90582-1} {\bibfield
  {journal} {\bibinfo  {journal} {Nucl.Phys.}\ }\textbf {\bibinfo {volume}
  {B275}},\ \bibinfo {pages} {734} (\bibinfo {year} {1986})}\BibitemShut
  {NoStop}%
%%CITATION = NUPHA,B275,734;%%
\bibitem [{\citenamefont {Philipsen}(2010)}]{Philipsen:2010gj}%
  \BibitemOpen
  \bibfield  {author} {\bibinfo {author} {\bibfnamefont {O.}~\bibnamefont
  {Philipsen}},\ }\href@noop {} {\  (\bibinfo {year} {2010})},\ \Eprint
  {http://arxiv.org/abs/1009.4089} {arXiv:1009.4089 [hep-lat]} \BibitemShut
  {NoStop}%
%%CITATION = ARXIV:1009.4089;%%
\bibitem [{\citenamefont {de~Forcrand}\ and\ \citenamefont
  {Philipsen}(2002)}]{deForcrand:2002ci}%
  \BibitemOpen
  \bibfield  {author} {\bibinfo {author} {\bibfnamefont {P.}~\bibnamefont
  {de~Forcrand}}\ and\ \bibinfo {author} {\bibfnamefont {O.}~\bibnamefont
  {Philipsen}},\ }\href {\doibase 10.1016/S0550-3213(02)00626-0} {\bibfield
  {journal} {\bibinfo  {journal} {Nucl.Phys.}\ }\textbf {\bibinfo {volume}
  {B642}},\ \bibinfo {pages} {290} (\bibinfo {year} {2002})},\ \Eprint
  {http://arxiv.org/abs/hep-lat/0205016} {arXiv:hep-lat/0205016 [hep-lat]}
  \BibitemShut {NoStop}%
%%CITATION = HEP-LAT/0205016;%%
\bibitem [{\citenamefont {D'Elia}\ and\ \citenamefont
  {Lombardo}(2003)}]{D'Elia:2002gd}%
  \BibitemOpen
  \bibfield  {author} {\bibinfo {author} {\bibfnamefont {M.}~\bibnamefont
  {D'Elia}}\ and\ \bibinfo {author} {\bibfnamefont {M.-P.}\ \bibnamefont
  {Lombardo}},\ }\href {\doibase 10.1103/PhysRevD.67.014505} {\bibfield
  {journal} {\bibinfo  {journal} {Phys.Rev.}\ }\textbf {\bibinfo {volume}
  {D67}},\ \bibinfo {pages} {014505} (\bibinfo {year} {2003})},\ \Eprint
  {http://arxiv.org/abs/hep-lat/0209146} {arXiv:hep-lat/0209146 [hep-lat]}
  \BibitemShut {NoStop}%
%%CITATION = HEP-LAT/0209146;%%
\bibitem [{\citenamefont {Bonati}\ \emph {et~al.}(2013)\citenamefont {Bonati},
  \citenamefont {D'Elia}, \citenamefont {de~Forcrand}, \citenamefont
  {Philipsen},\ and\ \citenamefont {Sanfillippo}}]{Bonati:2013tqa}%
  \BibitemOpen
  \bibfield  {author} {\bibinfo {author} {\bibfnamefont {C.}~\bibnamefont
  {Bonati}}, \bibinfo {author} {\bibfnamefont {M.}~\bibnamefont {D'Elia}},
  \bibinfo {author} {\bibfnamefont {P.}~\bibnamefont {de~Forcrand}}, \bibinfo
  {author} {\bibfnamefont {O.}~\bibnamefont {Philipsen}}, \ and\ \bibinfo
  {author} {\bibfnamefont {F.}~\bibnamefont {Sanfillippo}},\ }\href@noop {} {\
  (\bibinfo {year} {2013})},\ \Eprint {http://arxiv.org/abs/1311.0473}
  {arXiv:1311.0473 [hep-lat]} \BibitemShut {NoStop}%
%%CITATION = ARXIV:1311.0473;%%
\bibitem [{\citenamefont {Binder}(1981)}]{Binder:1981sa}%
  \BibitemOpen
  \bibfield  {author} {\bibinfo {author} {\bibfnamefont {K.}~\bibnamefont
  {Binder}},\ }\href {\doibase 10.1007/BF01293604} {\bibfield  {journal}
  {\bibinfo  {journal} {Z.Phys.}\ }\textbf {\bibinfo {volume} {B43}},\ \bibinfo
  {pages} {119} (\bibinfo {year} {1981})}\BibitemShut {NoStop}%
%%CITATION = ZEPYA,B43,119;%%
\bibitem [{Note1()}]{Note1}%
  \BibitemOpen
  \bibinfo {note} {Strictly speaking, critical exponents can be defined for
  second-order transitions only. However, similar considerations can be carried
  out for the first-order case, too, cf. \cite {Fisher:1982xt}.}\BibitemShut
  {Stop}%
\bibitem [{\citenamefont {{Khronos Working Group}}()}]{opencl}%
  \BibitemOpen
  \bibfield  {author} {\bibinfo {author} {\bibnamefont {{Khronos Working
  Group}}},\ }\href@noop {} {\enquote {\bibinfo {title} {The {OpenCL}
  {Specification}},}\ }\bibinfo {note}
  {Http://www.khronos.org/registry/cl/}\BibitemShut {NoStop}%
\bibitem [{\citenamefont {Bach}\ \emph
  {et~al.}(2013{\natexlab{a}})\citenamefont {Bach}, \citenamefont
  {Lindenstruth}, \citenamefont {Philipsen},\ and\ \citenamefont
  {Pinke}}]{Bach:2012iw}%
  \BibitemOpen
  \bibfield  {author} {\bibinfo {author} {\bibfnamefont {M.}~\bibnamefont
  {Bach}}, \bibinfo {author} {\bibfnamefont {V.}~\bibnamefont {Lindenstruth}},
  \bibinfo {author} {\bibfnamefont {O.}~\bibnamefont {Philipsen}}, \ and\
  \bibinfo {author} {\bibfnamefont {C.}~\bibnamefont {Pinke}},\ }\href
  {\doibase 10.1016/j.cpc.2013.03.020} {\bibfield  {journal} {\bibinfo
  {journal} {Comput.Phys.Commun.}\ }\textbf {\bibinfo {volume} {184}},\
  \bibinfo {pages} {2042} (\bibinfo {year} {2013}{\natexlab{a}})},\ \Eprint
  {http://arxiv.org/abs/1209.5942} {arXiv:1209.5942 [hep-lat]} \BibitemShut
  {NoStop}%
%%CITATION = ARXIV:1209.5942;%%
\bibitem [{\citenamefont {Bach}\ \emph {et~al.}(2011)\citenamefont {Bach},
  \citenamefont {Kretz}, \citenamefont {Lindenstruth},\ and\ \citenamefont
  {Rohr}}]{Bach2011a}%
  \BibitemOpen
  \bibfield  {author} {\bibinfo {author} {\bibfnamefont {M.}~\bibnamefont
  {Bach}}, \bibinfo {author} {\bibfnamefont {M.}~\bibnamefont {Kretz}},
  \bibinfo {author} {\bibfnamefont {V.}~\bibnamefont {Lindenstruth}}, \ and\
  \bibinfo {author} {\bibfnamefont {D.}~\bibnamefont {Rohr}},\ }\href {\doibase
  10.1007/s00450-011-0161-5} {\bibfield  {journal} {\bibinfo  {journal}
  {Computer Science - Research and Development}\ ,\ \bibinfo {pages} {1}}
  (\bibinfo {year} {2011})}\BibitemShut {NoStop}%
\bibitem [{\citenamefont {Bach}\ \emph
  {et~al.}(2013{\natexlab{b}})\citenamefont {Bach}, \citenamefont {Philipsen},\
  and\ \citenamefont {Pinke}}]{sanam}%
  \BibitemOpen
  \bibfield  {author} {\bibinfo {author} {\bibfnamefont {M.}~\bibnamefont
  {Bach}}, \bibinfo {author} {\bibfnamefont {O.}~\bibnamefont {Philipsen}}, \
  and\ \bibinfo {author} {\bibfnamefont {C.}~\bibnamefont {Pinke}},\
  }\href@noop {} {\bibfield  {journal} {\bibinfo  {journal} {PoS}\ }\textbf
  {\bibinfo {volume} {LAT2013}} (\bibinfo {year}
  {2013}{\natexlab{b}})}\BibitemShut {NoStop}%
\bibitem [{\citenamefont {Ferrenberg}\ and\ \citenamefont
  {Swendsen}(1989)}]{Ferrenberg:1989ui}%
  \BibitemOpen
  \bibfield  {author} {\bibinfo {author} {\bibfnamefont {A.~M.}\ \bibnamefont
  {Ferrenberg}}\ and\ \bibinfo {author} {\bibfnamefont {R.~H.}\ \bibnamefont
  {Swendsen}},\ }\href {\doibase 10.1103/PhysRevLett.63.1195} {\bibfield
  {journal} {\bibinfo  {journal} {Phys.Rev.Lett.}\ }\textbf {\bibinfo {volume}
  {63}},\ \bibinfo {pages} {1195} (\bibinfo {year} {1989})}\BibitemShut
  {NoStop}%
%%CITATION = PRLTA,63,1195;%%
\bibitem [{\citenamefont {Pelissetto}\ and\ \citenamefont
  {Vicari}(2002)}]{Pelissetto:2000ek}%
  \BibitemOpen
  \bibfield  {author} {\bibinfo {author} {\bibfnamefont {A.}~\bibnamefont
  {Pelissetto}}\ and\ \bibinfo {author} {\bibfnamefont {E.}~\bibnamefont
  {Vicari}},\ }\href {\doibase 10.1016/S0370-1573(02)00219-3} {\bibfield
  {journal} {\bibinfo  {journal} {Phys.Rept.}\ }\textbf {\bibinfo {volume}
  {368}},\ \bibinfo {pages} {549} (\bibinfo {year} {2002})},\ \Eprint
  {http://arxiv.org/abs/cond-mat/0012164} {arXiv:cond-mat/0012164 [cond-mat]}
  \BibitemShut {NoStop}%
%%CITATION = COND-MAT/0012164;%%
\bibitem [{Note2()}]{Note2}%
  \BibitemOpen
  \bibinfo {note}
  {Http://www.fz-juelich.de/ias/jsc/EN/Expertise/\\Supercomputers/JUQUEEN/JUQUEEN\protect
  \_node.html}\BibitemShut {NoStop}%
\bibitem [{Note3()}]{Note3}%
  \BibitemOpen
  \bibinfo {note} {We thank Georg Bergner for help with this
  analysis.}\BibitemShut {Stop}%
\bibitem [{\citenamefont {Fisher}\ and\ \citenamefont
  {Berker}(1982)}]{Fisher:1982xt}%
  \BibitemOpen
  \bibfield  {author} {\bibinfo {author} {\bibfnamefont {M.}~\bibnamefont
  {Fisher}}\ and\ \bibinfo {author} {\bibfnamefont {A.}~\bibnamefont
  {Berker}},\ }\href {\doibase 10.1103/PhysRevB.26.2507} {\bibfield  {journal}
  {\bibinfo  {journal} {Phys.Rev.}\ }\textbf {\bibinfo {volume} {B26}},\
  \bibinfo {pages} {2507} (\bibinfo {year} {1982})}\BibitemShut {NoStop}%
%%CITATION = PHRVA,B26,2507;%%
\end{thebibliography}%

\newpage
\onecolumngrid
% \clearpage

\appendix*
\section{Simulation details}
\label{ch:appendix}

In this section, details about the results carried out in the setup described in Section \ref{ch:results_wilson} will be given.
An overview about the simulated systems can be seen in Table \ref{tab:muiSims}.
Analysis details are given in Tables \ref{tab:muiSims_betaCPolyIm},\ref{tab:muiSims_betaCPbp} and \ref{tab:muiSims_betaCDeconf}.
The results of the fits of the Binder cumulant to \eqref{BinderScaling} are given in Table \ref{tab:muiSims_fitResults}.

\onecolumngrid
% \newpage
\vfill
\begin{table}[h!]
\centering
\begin{tabular}{|c|c|c|c|c|c|c|c|}
	\hline
	\LatMassWilson & \LatCoupling-range	& \NSigma=8	& \NSigma=10	& \NSigma=12	&  \NSigma=14	&  \NSigma=16	&   \NSigma=20	 \\
	\hline
	\hline
	0.1525 & 5.407-5.417 & 5.4120(6)	& 5.4132(3) & 5.4128(2) & 5.4128(2) & 5.4126(1) 	& -	\\
	0.1550 & 5.380-5.389 & -						& 5.3851(4) & - 					& 5.3852(2) & - 					& -	\\
	0.1575 & 5.350-5.361 & 5.3552(5) 	& 5.3553(3) & 5.3554(2) & 5.3555(1) & 5.3553(1) 	& -	\\
	0.1600 & 5.319-5.330 & 5.3220(4) 	& 5.3229(3) & 5.3226(2) & 5.3229(1) & 5.3231(2) 	& -	\\
	0.1625 & 5.284-5.294 & 5.2867(4) 	& 5.2875(2) & 5.2879(2) & 5.2881(1) & 5.2882(2)	& -	\\
	0.1650 & 5.246-5.256 & 5.2488(3) 	& 5.2510(2) & 5.2508(2)	& 5.2504(1) & 5.2510(1) & -	\\
\hline
\end{tabular} 
\caption{
		Overview of \LatCouplingC\ obtained from the peak of the susceptibility of \chiralcond\ .
	}
	\label{tab:muiSims_betaCPbp}
\end{table} 
\vfill

\begin{table}
\centering
\begin{tabular}{|c|c|c|c|c|c|c|c|}
	\hline
	\LatMassWilson & \LatCoupling-range	& \NSigma=8	& \NSigma=10	& \NSigma=12	&  \NSigma=14	&  \NSigma=16	&   \NSigma=20	 \\
	\hline
	\hline
	0.0300 & 5.685-5.696 & - 		& 40k	& 40k & 40k & 40k & -\\
	0.0400 & 5.685-5.695 & - 		& 40k	& 40k & 40k & 40k & -\\
	0.0500 & 5.683-5.695 & - 		& 40k	& 40k & 40k & 40k & -\\
	0.0600 & 5.681-5.695 & 40k	& 40k	& 40k & 40k & 40k & (40k)\\
	0.0650 & 5.676-5.689 & 40k 	& 40k	& 80k & -		 & 40k & -\\
	0.0700 & 5.676-5.688 & 40k 	& -		& 40k & -		 & 60k & 80k\\
	0.0865 & 5.662-5.678 & 40k 	& - 	& 40k & -		 & 60k & (80k)\\
	0.0910 & 5.659-5.673 & 40k 	& -		& 40k & -		 & 40k & 80k\\
	0.1000 & 5.647-5.658 & 40k 	& -		& 40k & -		 & 40k & 80k\\
	0.1040 & 5.640-5.655 & 40k	&	- 	& 40k & -		 & 40k & -\\
	0.1050 & 5.638-5.650 & 40k	&	- 	& 40k & -		 & 40k & -\\
	0.1060 & 5.638-5.650 & 40k	&	- 	& 40k & -		 & 40k & -\\
	0.1100 & 5.629-5.640 & 40k	&	- 	& 40k & -		 & 40k & 80k\\
	0.1200 & 5.602-5.613 & 40k	&	- 	& 40k & -		 & 40k & 80k\\
	0.1300 & 5.562-5.578 & 40k	&	- 	& 40k & -		 & 40k & -\\
	0.1400 & 5.508-5.520 & 40k	&	- 	& 40k & -		 & 40k & -\\
	0.1450 & 5.474-5.485 & 40k	&	- 	& 40k & 40k & 40k & -\\
	0.1500 & 5.431-5.441 & 40k	&	- 	& 40k & -		 & 40k & 80k\\
	0.1525 & 5.407-5.417 & 40k	&	40k & 40k & 40k & 40k & -\\
	0.1550 & 5.380-5.389 & 40k	&	40k & 40k & 40k & 40k & -\\
	0.1575 & 5.350-5.361 & 40k	&	40k & 40k & 40k & 40k & -\\
	0.1600 & 5.319-5.330 & 40k	&	40k & 40k & 40k & 40k & -\\
	0.1625 & 5.284-5.294 & 40k	&	40k & 40k & 40k & 40k & -\\
	0.1650 & 5.246-5.256 & 40k	&	40k & 40k & 40k & 40k & -\\
	\hline
\end{tabular} 
\caption{
		Overview of simulations carried out at $\MuI = \ii\pi\Temp$ and $\NTau=4$.
		The numbers given denote the statistics produced on each \LatCoupling\ point.
		A given \LatCoupling-range was scanned with $\Delta\LatCoupling=0.001$ for each \NSigma.
		Numbers in brackets indicate that some \LatCoupling\ values have smaller statistics.
	}
	\label{tab:muiSims}
\end{table}

\begin{table}
\centering
\begin{tabular}{|c|c|c|c|c|c|c|c|}
	\hline
	\LatMassWilson & \LatCoupling-range	& \NSigma=8	& \NSigma=10	& \NSigma=12	&  \NSigma=14	&  \NSigma=16	&   \NSigma=20	 \\
	\hline
	\hline
	0.0300 & 5.685-5.696 & - 					& 5.6903(6)	& 5.6905(4) 	& 5.6906(3) 	& 5.6915(2) 	& -\\
	0.0400 & 5.685-5.695 & - 					& 5.6903(4)	& 5.6910(4) 	& 5.6906(3) 	& 5.6904(2) 	& -\\
	0.0500 & 5.683-5.695 & - 					& 5.6892(4)	& 5.6898(3)		& 5.6904(3)		& 5.6897(2)		& -\\
	0.0600 & 5.681-5.695 & 5.6878(4)	& 5.6873(5)	& 5.6883(4) 	& 5.6881(3) 	& 5.6873(2) 	& -\\
	0.0650 & 5.676-5.689 & 5.6869(6) 	& 5.6857(4)	& 5.6856(2)  & - 					& 5.6855(2) 	& -\\
	0.0700 & 5.676-5.688 & 5.6856(5)	& - 				& 5.6844(4) 	& - 					& 5.6833(2) 	& 5.6832(2) \\
	0.0865 & 5.662-5.678 & 5.6745(5)	& - 				& 5.6723(3) 	& - 					& 5.6714(2) 	& 5.6716(1) \\
	0.0910 & 5.659-5.673 & 5.6705(5) & - 					& 5.6673(4) 	& - 					& 5.6668(2) 	& 5.6666(2) \\
	0.1000 & 5.647-5.658 & 5.6565(6) & - 					& 5.6550(3) 	& - 					& 5.6545(3) 	& 5.6547(1) \\
	0.1040 & 5.640-5.655 & 5.6498(6) & - 					& 5.6489(3) 	& - 					& 5.6481(3) 	& - \\
	0.1050 & 5.638-5.650 & 5.6488(5) & - 					& 5.6469(3) 	& - 					& 5.6460(3) 	& - \\
	0.1060 & 5.638-5.650 & 5.6471(5) & - 					& 5.6448(3) 	& - 					& 5.6444(3) 	& - \\
	0.1100 & 5.629-5.640 & 5.6385(6) & - 					& 5.6365(4) 	& - 					& 5.6360(3) 	& 5.6360(2) \\
	0.1200 & 5.602-5.613 & 5.6113(5) & - 					& 5.6104(3) 	& - 					& 5.6093(2) 	& 5.6092(2) \\
	0.1300 & 5.562-5.578 & 5.5743(5) & - 					& 5.5725(2) 	& - 					& 5.5715(2) 	& - \\
	0.1400 & 5.508-5.520 & 5.5193(4) & - 					& 5.5182(3) 	& - 					& 5.5166(3) 	& - \\
	0.1450 & 5.474-5.485 & 5.4828(4) & - 					& 5.4817(3) 	& 5.4812(2) 	& 5.4807(2) 	& - \\
	0.1500 & 5.431-5.441 & 5.4384(4) & - 					& 5.4377(3) 	& - 					& 5.4376(2)		& 5.4373(1) \\
	0.1525 & 5.407-5.417 & 5.4137(3) & 5.4131(3) & 5.4129(2) & 5.4127(2) 	& 5.4124(2) 	& - \\
	0.1550 & 5.380-5.389 & 5.3854(3) & 5.3852(3) & 5.3855(3) & 5.3852(2) 	& 5.3851(1) 	& - \\
	0.1575 & 5.350-5.361 & 5.3560(3) & 5.3555(3) & 5.3555(2) & 5.3555(1) 	& 5.3554(2) 	& - \\
	0.1600 & 5.319-5.330 & 5.3230(3) & 5.3237(2) & 5.3227(2) & 5.3229(1) 	& 5.3231(1) 	& - \\
	0.1625 & 5.284-5.294 & 5.2876(3) & 5.2877(2) & 5.2880(2) & 5.2881(1) 	& 5.2880(1) 	& - \\
	0.1650 & 5.246-5.256 & 5.2493(3) & 5.2510(2) & 5.2509(2) & 5.2504(1) 	& 5.2510(1) 	& - \\
	\hline
\end{tabular} 
\caption{
		Overview of \LatCouplingC\ obtained from the peak of the susceptibility of \PolyAbs\ .
	}
	\label{tab:muiSims_betaCDeconf}
\end{table} 

\begin{table}
\centering
\begin{tabular}{|c|c|c|c|c|c|c|c|}
	\hline
	\LatMassWilson & \LatCoupling-range	& \NSigma=8	& \NSigma=10	& \NSigma=12	&  \NSigma=14	&  \NSigma=16	&   \NSigma=20	 \\
	\hline
	\hline
	0.0300 & 5.685-5.696 & - 					& - 					& - 					& - 					& - 					& -						\\
	0.0400 & 5.685-5.695 & - 					& - 					& - 					& - 					& 5.6920(4) 	& -						\\
	0.0500 & 5.683-5.695 & - 					& - 					& - 					& 5.6928(6) 	& 5.6914(5) 	& -						\\
	0.0600 & 5.681-5.695 & - 					& - 					& 5.6897(5) 	& 5.6888(3) 	& 5.6873(3) 	& -						\\
	0.0650 & 5.676-5.689 & - 					& - 					& 5.6862(3) 	& - 					& 5.6854(2) 	& -						\\
	0.0700 & 5.676-5.688 & - 					& - 					& 5.6847(4) 	& - 					& 5.6832(2) 	& 5.6831(2)		\\
	0.0865 & 5.662-5.678 & 5.6765(7) 	& - 					& 5.6720(3) 	& - 					& 5.6712(2) 	& 5.6715(1)		\\
	0.0910 & 5.659-5.673 & 5.6707(6) 	& - 					& 5.6671(4) 	& - 					& 5.6666(2) 	& 5.6665(2)		\\
	0.1000 & 5.647-5.658 & 5.6561(6) 	& - 					& 5.6546(3) 	& - 					& 5.6541(3) 	& 5.6543(2)		\\
	0.1040 & 5.640-5.655 & 5.6493(6) 	& - 					& 5.6484(3) 	& - 					& 5.6476(2) 	& - 					\\
	0.1050 & 5.638-5.650 & 5.6483(5)	& - 					& 5.6464(3) 	& - 					& 5.6455(2) 	& -						\\
	0.1060 & 5.638-5.650 & 5.6465(5)	& - 					& 5.6442(3) 	& - 					& 5.6439(3) 	& - 					\\
	0.1100 & 5.629-5.640 & 5.6378(6) 	& - 					& 5.6358(4) 	& - 					& 5.6354(3) 	& 5.6354(2)		\\
	0.1200 & 5.602-5.613 & 5.6106(5) 	& - 					& 5.6094(3) 	& - 					& 5.6085(3) 	& 5.6085(2)		\\
	0.1300 & 5.562-5.578 & 5.5734(5) 	& - 					& 5.5713(2)	 	& - 					& 5.5706(2) 	& - 					\\
	0.1400 & 5.508-5.520 & 5.5181(4) 	& - 					& 5.5169(3) 	& - 					& 5.5155(3) 	& - 					\\
	0.1450 & 5.474-5.485 & 5.4815(4) 	& - 					& 5.4806(3) 	& 5.4803(2) 	& 5.4799(2) 	& - 					\\
	0.1500 & 5.431-5.441 & 5.4372(4) 	& - 					& 5.4367(2) 	& - 					& 5.4369(1) 	& 5.4368(1)		\\
	0.1525 & 5.407-5.417 & 5.4127(3) 	& 5.4122(3) 	& 5.4121(2) 	& 5.4120(2) 	& 5.4118(2)		& -						\\
	0.1550 & 5.380-5.389 & 5.3845(3) 	& 5.3844(3) 	& 5.3848(3) 	& 5.3847(2) 	& 5.3847(1) 	& -						\\
	0.1575 & 5.350-5.361 & 5.3553(3) 	& 5.3549(3) 	& 5.3550(2) 	& 5.3551(2) 	& 5.3551(2) 	& -						\\
	0.1600 & 5.319-5.330 & 5.3224(3) 	& 5.3232(3) 	& 5.3224(2) 	& 5.3226(1) 	& 5.3229(2) 	& -						\\
	0.1625 & 5.284-5.294 & 5.2872(3) 	& 5.2874(2) 	& 5.2878(2) 	& 5.2880(1) 	& 5.2879(1) 	& -						\\
	0.1650 & 5.246-5.256 & 5.2490(3) 	& 5.2509(2) 	& 5.2508(2) 	& 5.2504(1) 	& 5.2510(1) 	& -						\\
\hline
\end{tabular} 
\caption{
		Overview of \LatCouplingC\ obtained from the peak of the susceptibility of $|\PolyIm|$.
	}
	\label{tab:muiSims_betaCPolyIm}
\end{table} 
\begin{table}
\centering
\begin{tabular}{|c|c|c|c|c|c|c|c|}
	\hline
	\LatMassWilson & \NSigma	& \LatCouplingC	& $\nu$	& $\Binder(\LatCoupling, \infty)$ & $a_1$	&  $a_2$	&   $\chi^2$ \\
	\hline
	\hline
	0.0300 & 12,14,16 	& 5.6921(3) &   0.289(57)  &  2.37(4) &   -0.016(29) & - &   0.968\\
	0.0400 & 12,14,16 	& 5.6883(2) & 0.305(70) &  2.63(4) &  -0.036(73)	& -		& 0.983 \\
	0.0500 & 10,12,14,16 	& 5.6891(1) & 0.351(34) &  2.18(2) &  -0.089(66)	& - 		& 0.965 \\	
	0.0600 & 10,12,14,16 	& 5.6862(1) & 0.369(14) &  2.04(1) &  -0.126(50)	& 0.005(3) 	& 0.773 \\	
	0.0650 & 10,12,16 	& 5.6844(1) & 0.375(13) &  1.92(1) &  -0.13(30)	& 0.007(3) 	& 1.408 \\
	0.0700 & 12,16,20 	& 5.6829(1) & 0.423(39) &  1.86(2) &  -0.28(18)	& - 		& 1.029 \\
	0.0865 & 8,12,16 		& 5.6704(1) & 0.450(18) &  1.89(1) &  -0.378(09)	& - 		& 1.026 \\
	0.0910 & 8,12,16,20 	& 5.6655(1) & 0.458(22) &  1.85(1) &  -0.38(11)  	& 0.062(34) 	& 1.173 \\
	0.1000 & 8,12,16,20 	& 5.6539(1) & 0.501(19) &  1.74(1) &  -0.56(12)	& -		& 0.952 \\
	0.1040 & 8,12,16 		& 5.6469(1) & 0.547(17) &  1.77(1) &  -0.79(11) 	& - 		& 0.991 \\
	0.1050 & 8,12,16		& 5.6438(1) & 0.650(23) &  1.85(1) &  -1.52(21)	& - 		& 1.019 \\
	0.1060 & 8,12,16		& 5.6425(1) & 0.589(23) &  1.82(1) &  -1.10(18)	& -		& 1.015 \\
	0.1100 & 8,12,16, 20	& 5.6341(1) & 0.582(43) &  1.80(1) &  -1.08(38) 	& - 		& 1.064 \\ 
	0.1200 & 12,16,20		& 5.6075(1) & 0.598(21) &  1.75(1) &  -1.167(20)	& 0.40(14)	& 0.996 \\
	0.1300 & 8, 12, 16	& 5.5689(1) & 0.637(22) &  1.83(1) &  -1.64(22) 	& - 		& 0.860 \\
	0.1400 & 8, 12, 16	& 5.5146(1) & 0.612(20) &  1.83(1) &  -1.59(21) 	& -		& 0.821 \\ 
	0.1450 & 8, 12, 16	& 5.4790(1) & 0.588(23) &  1.80(1) &  -1.50(26) 	& - 		& 1.003 \\
	0.1500 & 12, 16, 20	& 5.4367(1) & 0.611(42) &  1.66(2) &  -2.12(69) 	& -		& 0.950 \\ 
	0.1525 & 10, 12, 14, 16 	& 5.4114(1) & 0.620(39) &  1.76(2) &  -2.51(66)	& - 		& 1.029 \\ 
	0.1550 & 10, 12, 14, 16 	& 5.3849(1) & 0.512(37) &  1.67(1) &  -1.26(46) 	& -		& 0.984\\ 
	0.1575 & 12, 14, 16  	& 5.3548(1) & 0.555(28) &  1.80(2) &  -2.42(60) 	& 2.15(1.05)	& 1.015 \\
	0.1600 & 8, 10, 12 	& 5.3225(1) & 0.376(37) &  1.77(2) &  -0.43(26)	& -		& 0.997 \\
	0.1625 & 10, 12, 14 	& 5.2886(1) & 0.331(18) &  1.58(1) &  -0.16(07)	& 0.015(12)	& 0.983 \\
	0.1650 & 10, 12, 14 	& 5.2501(1) & 0.364(47) &  2.15(6) &  -0.67(61) 	& 0.11(20)	& 0.981 \\
	\hline
\end{tabular} 
\caption{
		Overview of fits to $\Binder(\PolyIm)$ according to \eqref{BinderScaling}.
		The \NSigma\ column indicates which datasets have been used in the fit, see also Table \ref{tab:muiSims}.
		If no value for $a_2$ is given, the fit has been performed with the linear ansatz.
	}
	\label{tab:muiSims_fitResults}
\end{table}

\end{document}